\documentclass[journal]{IEEEtran}

\usepackage{cite}

\usepackage{amsmath,amssymb,amssymb,fancyhdr,wasysym,graphicx}
\usepackage{graphicx} %added by me
\usepackage{multirow} %added by me
\usepackage{amsmath, amsfonts, amssymb}%added by me
\usepackage{url}%added by me
\usepackage{xcolor}%added by me
\usepackage{threeparttable}%added by me
\usepackage[perpage,symbol]{footmisc}
\usepackage{booktabs}
\usepackage{makecell}
\setfnsymbol{wiley}
\usepackage{CJK}% used to print circled text.

\ifCLASSINFOpdf
\else
\fi
\begin{document}

\title{Machine Learning Attack and Defense on Voltage Over-scaling-based Lightweight Authentication}

\author{Jiliang~Zhang,~\IEEEmembership{Member,~IEEE,} ~Haihan~Su, ~Gang~Qu,~\IEEEmembership{Senior Member,~IEEE}

%,~Zheng~Qin

\thanks{This work is supported by the National Natural Science Foundation of China (Grant NO. 61874042, 61602107), the Hu-Xiang Youth Talent Program (Grant No. 2018RS3041), the National Natural Science Foundation of Hunan Province, China (Grant No. 618JJ3072), the 2017 CCF-IFAA RESEARCH FUND, and the Fundamental Research Funds for the Central Universities.}

\thanks{J. Zhang and H. Su are with the College of Computer Science and Electronic Engineering, Hunan University, Changsha 410082, China (e-mail: zhangjiliang@hnu.edu.cn).}

\thanks{G. Qu is with the Department of Electrical and Computer Engineering, University of Maryland, College Park, MD 20742 USA (e-mail: gangqu@umd.edu).}

}

\maketitle

% As a general rule, do not put math, special symbols or citations
% in the abstract or keywords.

\begin{abstract}
It is a challenging task to deploy lightweight security protocols in resource-constrained IoT applications. A hardware-oriented lightweight authentication protocol based on device signature generated during voltage over-scaling (VOS) was recently proposed to address this issue. VOS-based authentication employs the computation unit such as adders to generate the process variation dependent error which is combined with secret keys to create a two-factor authentication protocol. In this paper, machine learning (ML)-based modeling attacks to break such authentication is presented. We also propose a \underline{c}hallenge \underline{s}elf-\underline{o}bfuscation \underline{s}tructure (CSoS) which employs previous challenges combined with keys or random numbers to obfuscate the current challenge for the VOS-based authentication to resist ML attacks. Experimental results show that ANN, RNN and CMA-ES can clone the challenge-response behavior of VOS-based authentication with up to 99.65$\%$ prediction accuracy, while the prediction accuracy is less than 51.2$\%$ after deploying our proposed ML resilient technique. In addition, our proposed CSoS also shows good obfuscation ability for strong PUFs. Experimental results show that the modeling accuracies are below 54\% when $10^6$ CRPs are collected to model the CSoS-based Arbiter PUF with ML attacks such as LR, SVM, ANN, RNN and CMA-ES.

\end{abstract}

% Note that keywords are not normally used for peerreview papers.
\section{Introduction}

%\subsection{Motivation}

The Internet of Things (IoT) is a novel networking paradigm which connects a variety of things or objects to the Internet through sensor technology, radio frequency identification (RFID), communication technology, computer networks and database technology \cite{IoT}. According to the IHS forecast \cite{IoT2017}, the IoT market will grow from an installed base of 15.4 billion devices in 2015 to 30.7 billion devices in 2020 and 75.4 billion in 2025. With the increasing of IoT devices, security issues have attracted much attention. For example, in 2016, America suffered the largest DDoS attack in history \cite{Woolf2016}. The cyber-attack that brought down much of America internet was caused by the Mirai botnet, which is a worm-like family of malware that infected IoT devices and corralled them into a DDoS botnet \cite{Antonakakis2017}. Therefore, secure and efficient defenses need to be deployed for IoT devices.

Secret key storage and device authentication are two key technologies for IoT security. Traditional key generation and authentication techniques are based on the classical cryptography, which requires expensive secret key storage and high-complexity cryptographic algorithms. In many IoT applications, resources like CPU, memory, and battery power are limited and cannot afford the classic cryptographic security solutions. Therefore, lightweight solutions for IoT security are urgent.

%\subsection{Limitations of Prior Art}

Physical unclonable functions (PUFs) \cite{Ruhrmair2014,Zhang2014} and recently proposed voltage over-scaling (VOS) based authentication \cite{Arafin2017} are two emerged lightweight security primitives for IoT device authentication.

\begin{figure}
\centerline{\includegraphics[width=\linewidth]{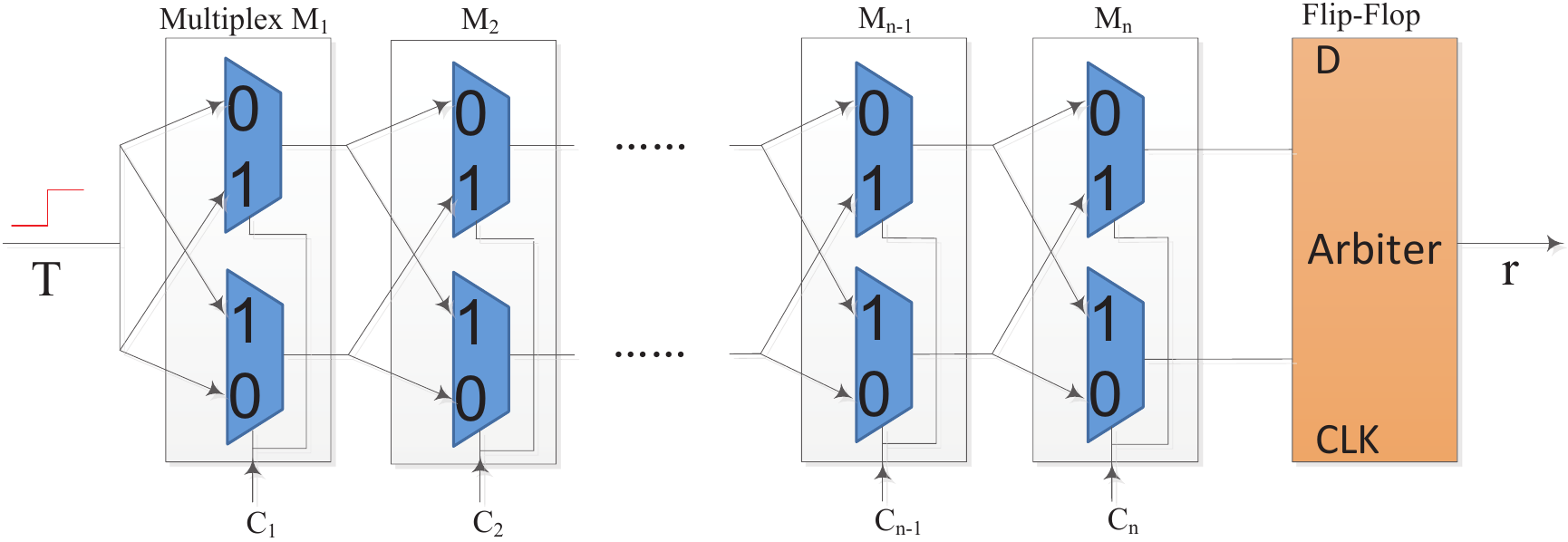}}
\caption{The structure of Arbiter PUF \cite{Lim2005}.}
\label{APUF}
\end{figure}

PUFs use a random factor caused by process variations in the manufacturing process to generate unclonable responses for challenges to authenticate devices. Since the PUF has been introduced \cite{Pappu2002}, it has attracted much attention as a low-cost alternative solution for key generation and device authentication, and hence many different PUF structures have been proposed.
PUFs can be broadly categorized into strong PUFs \cite{Lee2004,Lim2005,Vijayakumar2015,Majzoobi2008a,Sahoo2014} and weak PUFs \cite{Holcomb2007,Suh2007,Tuyls2006,Sauer2017}.
A weak PUF produces a small amount of stable CRPs that can be used as unique keys or seeds for traditional encryption systems. SRAM PUF \cite{Holcomb2007} and ring oscillator (RO) PUF \cite{Suh2007} are typical weak PUFs.
Arbiter PUF \cite{Lim2005} is a typical strong PUF, the circuit structure is shown in Fig. \ref{APUF}. Strong PUFs are based on their high entropy content and can provide a huge number of unique CRPs to authenticate the device.
However, the current strong PUFs are vulnerable to machine learning attacks that attackers can collect a certain number of CRPs to model the PUF easily.
For the traditional Arbiter PUF, a cloned model can be built with the prediction accuracy above 95$\%$ after only collecting 650 CRPs, which means the cloned model can exhibit the similar challenge-response behavior to the original PUF \cite{Zhang2018}.

Compared with the PUFs, the VOS-based authentication has two advantages \cite{Arafin2017}: 1) lower power consumption; 2) no additional hardware required. Therefore, the VOS-based authentication is more suitable for resource-constrained IoT applications. VOS is a common power reduction technology and can be used for approximate computing \cite{Venkatesan2011}. The calculation unit of digital circuits can generate correct results for all inputs under the normal operating voltage, but calculation errors may occur in VOS \cite{Chen2013}. Meanwhile, the errors generated by the computing unit in VOS are related to the manufacturing process variation and hence can be used as hardware fingerprints for device authentication. Recently, Arafin, Gao and Qu \cite{Arafin2017} proposed to use such errors generated by the computing unit in VOS as the device signatures and designed a two-factor authentication protocol, named \underline{v}oltage \underline{o}ver-scaling-based \underline{l}ightweight \underline{a}uthentication (VOLtA).

This paper proves that the VOLtA is vulnerable to machine learning (ML) attacks. We first report the ML attacks on VOLtA in \cite{Zhang2018}. In this article, 1) we elaborate the details of ML attacks on VOLtA; 2) In order to resist ML attacks, a \underline{c}hallenge \underline{s}elf-\underline{o}bfuscation \underline{s}tructure (CSoS) is new proposed against ML attacks for VOLtA, and it is a general obfuscation method that also can be used to secure Strong PUFs; 3) we verify the effectiveness of proposed ML attacks and defense by HSpice platform using the FreePDK 45nm libraries. The main contributions of this paper are as follows.

\begin{enumerate}
  \item %\textbf{\emph{Attack.}}
  We reevaluate the security of VOLtA. For the first time, we demonstrate that ML attacks such as artificial neural network (ANN), recurrent neural network (RNN) and covariance matrix adaptation evolution strategy (CMA-ES) can break VOLtA successfully. Especially, the prediction accuracy of RNN is up to 99.65$\%$.
  \item %\textbf{\emph{Defense.}}
  We propose a CSoS-based ML resistant authentication protocol that reduces the prediction accuracy of modeling to less than 51.2$\%$.
  \item %\textbf{\emph{Eliminate weakness.}}
  The VOS-based two-factor authentication scheme requires a very long key to encrypt the output, which incurs unacceptable key storage overhead. Our proposed CSoS-based ML resistant authentication protocol eliminates such weakness.
  \item %\textbf{\emph{Universality.}}
  %CSoS不仅适用于VOLtA，同时也能在传统的Strong PUF部署，并展现出优异的抗建模成效。我们收集部署了CSoS的Arbiter PUF的 $10^6$ 对CRPs，并使用LR，SVM，ANN以及CMA-ES对其进行建模攻击，建模准确率依旧在54$\%$ 以下。
  CSoS is not only efficient for VOLtA, but also can be deployed for strong PUFs and exhibits good obfuscation ability. After deploying the CSoS, the modeling accuracy for a Arbiter PUF is below 54$\%$ with LR, SVM, ANN, RNN and CMA-ES when $10^6$ CRPs are collected.
  \item %\textbf{\emph{No effects on uniqueness and reliability.}}
  CSoS uses the previous challenges combined with keys or random numbers to obfuscate the current challenge without changing the structure of the authentication circuit such as VOS-adders and PUFs. Therefore, it will not affect the uniqueness and reliability.
\end{enumerate}

%\subsection{Outline of the Paper}

The rest of this paper is organized as follows. Section II introduces some related definitions, concepts and terminologies. Section III gives a detailed security analysis for VOLtA and ML attack methods. The CSoS-based ML attacks resistant authentication is elaborated in Section IV. The detailed experimental results are reported in Section V. Finally, we give the conclusion in Section VI.

\section{Preliminaries}
This section will introduce the principle of generating calculation errors in the VOS circuit and the ML algorithms which are used to model the VOLtA.

\subsection{Voltage Over-scaling}
In digital signal processing systems, the power consumption $P$ is given by:

\begin{equation}
P = C_L V_{dd}^2 f_s
\end{equation}

\noindent where $V_{dd}$ is the supply voltage; $C_L$ is the effective switching capacitance; $f_s$ is the clock frequency of circuit \cite{Chen2013}.
According to Eqn. (1), the power consumption $P$ decreases with the operating voltage $V_{dd}$. Some techniques employ this feature to reduce the power consumption of circuit, such as multiple supply voltages \cite{Rokhani2006}, variable voltage scaling \cite{Gutnik1997} and retiming technique \cite{Chabini2004}.
The circuit delay $\tau_d$ is given by:

\begin{equation}
\tau_d = \frac{C_L V_{dd}}{\beta (V_{dd} - V_t) ^ \alpha}
\end{equation}

\noindent where $\alpha$ is the velocity saturation index, $\beta$ is the gate trans-conductance and $V_t$ is the device threshold voltage \cite{Arafin2017}. We can see from the Eqn. (1) and (2) that power consumption will decrease quadratically and the delay will increase dramatically with the decreasing of supply voltage \cite{Liu2011}. With the correct timing constraints, the circuit produces correct outputs for all inputs. However, when the operating voltage is lowered, the timing violations may incur calculation errors. In the approximate computing, the computing unit performs high-bit calculations in the normal voltage and calculates low-bits in VOS to generate approximate results and significantly reduces the power consumption \cite{Venkatesan2011,Han2013b,Li2013}. Furthermore, the errors produced by the process variation are random and can be reproduced by the original device but difficult to be cloned. Therefore, the errors can be used as the hardware fingerprints to authenticate the devices.

\subsection{Computing Errors}

%We use the Ripple Carry Adder (RCA), a common calculation unit in the digital circuit, as an example to explain why the errors would be produced in VOS. The layout of RCA circuit is simple. As shown in Fig. \ref{Fig1}(a), we connect multiple full-adders to form a multi-bit adder. The disadvantage of RCA is that its high-bit calculations must wait for the completion of low-bit operations. Therefore, the signal transmission delay is relatively long in RCA. Fortunately, the errors as the device signatures in VOLtA are caused by the circuit delay. The obvious delay will have a significant contribution for the errors produced, while the long delay can meet the requirements better in RCA \cite{Arafin2017,Venkatesan2011}.

\begin{figure}
\centerline{{\includegraphics[width=\linewidth]{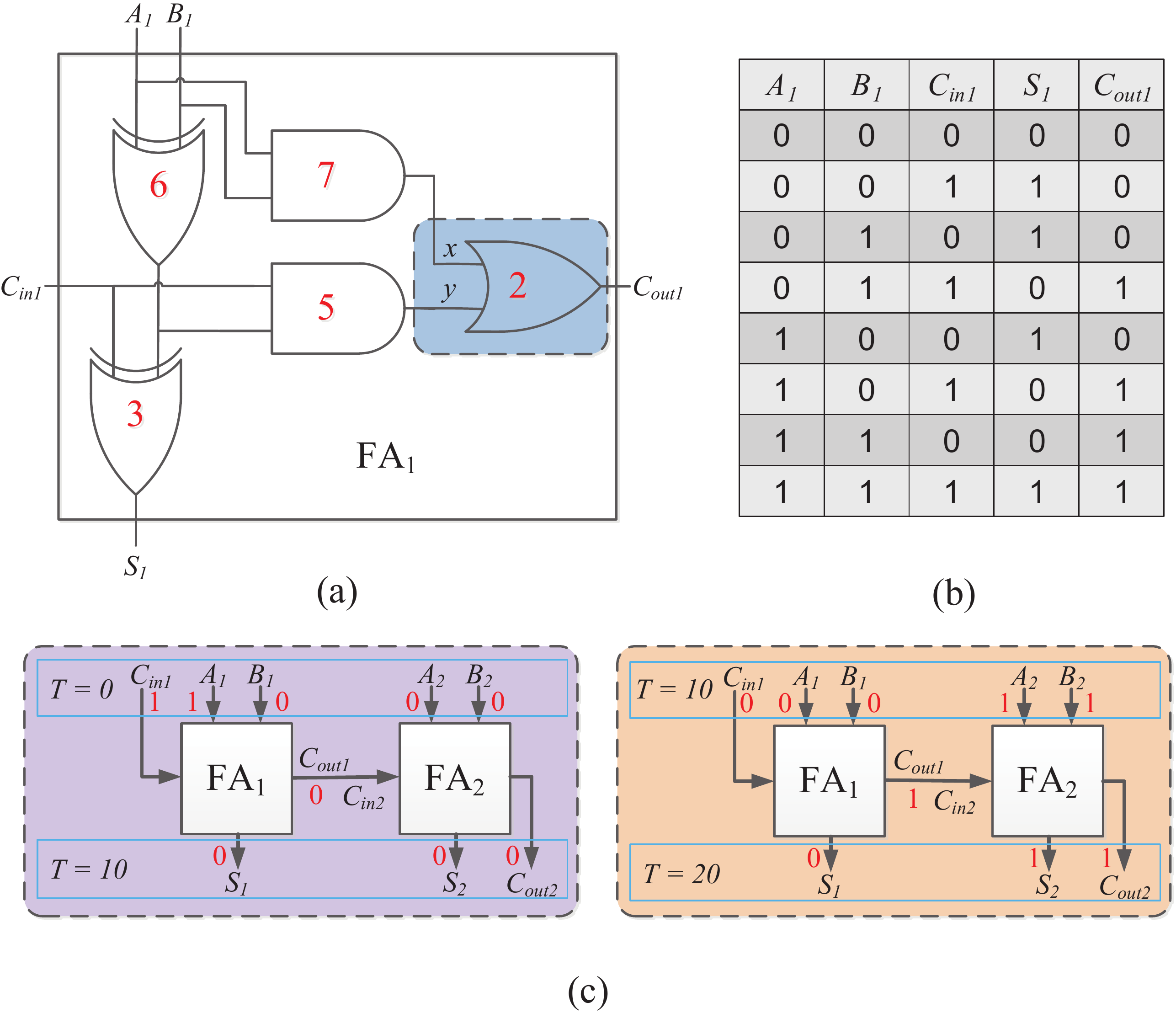}}}
\caption{An example of computing error. (a) The gate circuit of a full-adder. (b) The truth table of the full-adder. (c) The generation process of computing errors. A $n$-bit RCA is connected by $n$ full-adders, and 2-bit RCA is depicted in (c).}
\label{Fig1}
\end{figure}

 As a common computing unit in digital circuits, ripple carry adder (RCA) has the potential to preserve process variation related artifacts \cite{Arafin2017}. The principle of errors caused by the circuit delay is described in Fig. \ref{Fig1}. The gate circuit and truth table of a full-adder (FA) are shown in Fig. \ref{Fig1}(a) and Fig. \ref{Fig1}(b), respectively. Fig. \ref{Fig1}(c) gives the process of generating computing errors, where FA$_1$ is a simplified diagram of Fig. \ref{Fig1}(a). For ease of exposition, we assume that the red numbers marked in Fig. \ref{Fig1}(a) are the signal transmission delays of the logic gates, and there is no delay in FA$_2$. In Fig. \ref{Fig1}(c), when the clock period of the input signal is `10', the first clock period is as follows.

\begin{itemize}
\item At time $t = 0$, the input pulse signal $\{C_{in1}, A_1, B_1, A_2, B_2\} = \{1, 1, 0, 0, 0\}$;
\item At time $t = 10$, since the delay $D_y = 6 + 5 > 10$ at the $y$-input of OR gate, the signal '1' is not transmitted to $y$-input, hence the signal at $y$-input is still '0'. The $x$-input of OR gate delay $D_x = 7 < 10$, the signal '0' is transmitted to $x$-input successfully, and thus the $C_{out1}$-output of OR gate is '0'. The output $\{S_1, S_2, C_{out2}\} = \{0, 0, 0\} \neq \{0, 1, 0\}$, the first clock period is over.
\end{itemize}

\noindent The second clock period is as follows.

\begin{itemize}
\item At time $t = 10$, the input pulse signal $\{C_{in1}, A_1, B_1, A_2, B_2\} = \{0, 0, 0, 1, 1\}$;
\item At time $t = 20$, since $D_y = 6 + 5 < 20$, the signal `1' of the first clock period is transmitting in $C_{out1}$, and thus the output $\{S_1, S_2, C_{out2}\} = \{0, 1, 1\} \neq \{0, 0, 1\}$.
\end{itemize}

As discussed above, the errors produced by the adder in VOS are related to the current input and the previous inputs.

\subsection{ Machine Learning}

\begin{enumerate}
  \item [1)]\emph{Logistic Regression (LR)}
\end{enumerate}

In the device authentication, the response bit is '0' or '1', which is a binary classification problem. LR is a fast binary classification algorithm used in machine learning. As a binary classification model, logistic regression has multiple inputs, such as feature vector $X = (x_1, x_2, ..., x_n)$, and the output $Y$ is obtained by inputting $X$ into the classifier. The formula of the classifier is $Y=g(w_0+w_1x_1+w_2x_2+ ... +w_nx_n)$. Usually, LR uses the $sigmoid$ $g(z) = 1/(1 + e^{-z})$ to make $Y$ close to 0 or 1. Arbiter PUFs can be modeled by LR with the high accuracy \cite{Rührmair2010,Zhang2018}.

\begin{enumerate}
  \item [2)]\emph{Support Vector Machines (SVM)}
\end{enumerate}

SVM \cite{Bishop2015} can perform binary classification by mapping known training instances into a higher-dimensional space. The goal of SVM training is to find the most suitable separation hyperplane and solve the nonlinear classification tasks that cannot be linearly separated in the original space. The separation hyperplane should keep the maximum distance from all vectors of different classifications as much as possible. The vector with the smallest distance to the separation hyperplane is called the support vector. The separation hyperplane is constructed by the two parallel hyperplanes with support vectors of different classifications. The distance between the hyperplanes is called the margin. The key of constructing a good SVM is to maximize the margin while minimizing classification errors and the whole process is regulated by the regularization coefficient $\lambda$. In well-trained SVMs, kernel functions are often used to solve the problem of support vector selection and classification difficulties. There are three frequently-used kernel functions: 1) linear: $K(w,z) = z^T w $ (only solves linearly separable problems);  2) radial basis function (RBF): $K(w,z)$ = $\exp$$((-\|w-z||_2^2)/\sigma^2 )$; 3) multi-layer perception (MLP): $K(w,z) = \tanh$$(\alpha z^T w+\beta)$. Training a good SVM classifier always requires to adjust regularization coefficient $\lambda$, $\sigma^{2}$ (RBF) or $(\alpha,\beta)$ (MLP).

\begin{figure}
\centerline{\includegraphics[width=\linewidth]{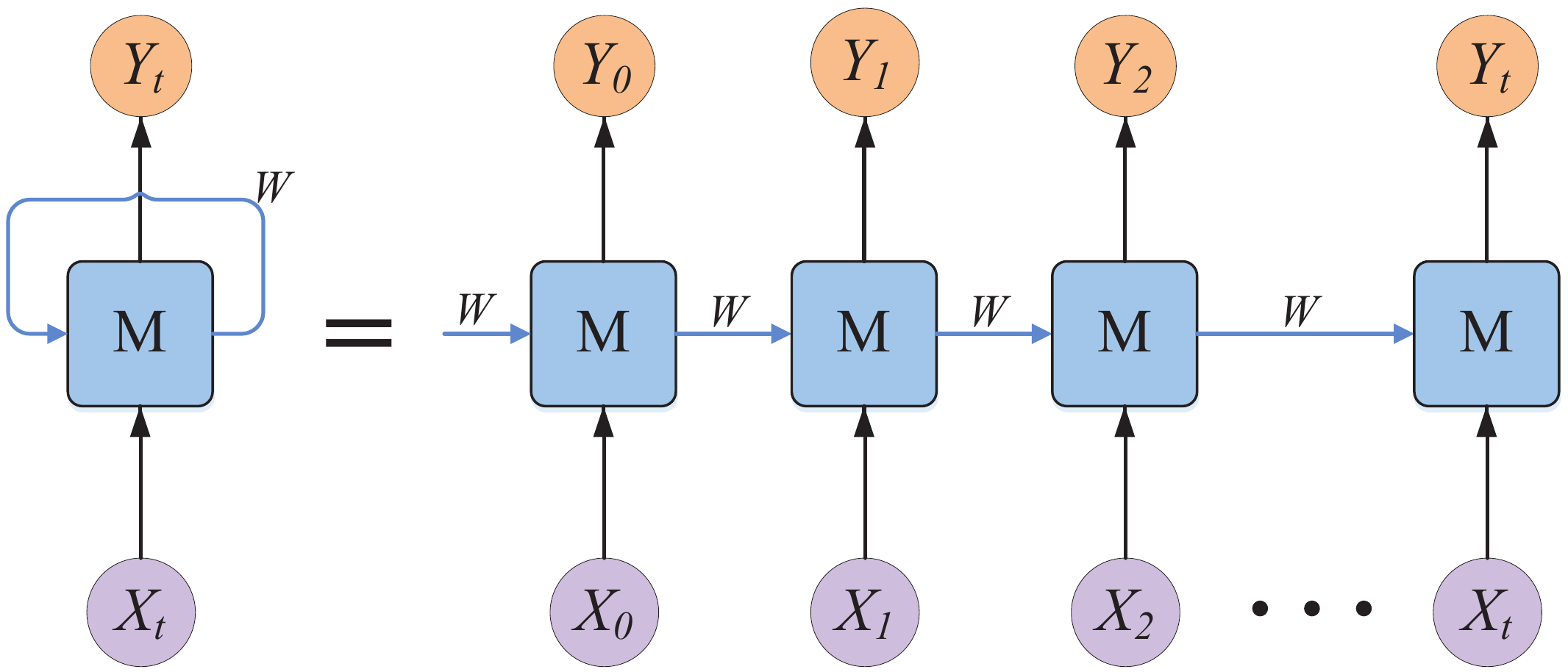}}
\caption{The structure of recurrent neural network}
\label{Fig2}
\end{figure}

\begin{enumerate}
  \item [3)]\emph{Artificial Neural Network (ANN)}
\end{enumerate}

ANN is interconnected by computational nodes called neurons, which has the adaptive capability. In other words, ANN can adjust the weight parameters utilizing the prepared training set to fit the required function. The universal approximation theorem \cite{Hornik1991} shows that if a pre-feedback neural network has a linear output layer and at least one hidden layer with an activation function such as $sigmoid$, it can fit any function with high accuracy as long as there are enough neurons. The simplest neural network comprises of a layer with several neurons, called a single layer perceptron (SLP) \cite{Rosenblatt1957}. For each neuron, all input vectors are weighted, added, biased, and applied to an activation function to generate an output. In the SLP training process, the neuron updates its weights and bias according to the linear feedback function of the training set prediction error. When the prediction accuracy or iterations of trained model reaches the predetermined value, the training process is terminated. This paper uses a simple 2-layer neural network structure to model the logic gates and the obfuscation mechanism with invariable key, and employs a 3-layer ANN (160 nodes in the first layer, 40 nodes in the second layer and 8 nodes in the third layer) to model VOLtA. In addition, we use $sigmoid$ as the activation function.

\begin{enumerate}
  \item [4)]\emph{Recurrent Neural Network (RNN)}
\end{enumerate}

RNN is mainly used to deal with sequence data. In the traditional neural network model, from the input layer through the hidden layer to the output layer, the layers are fully connected and the nodes in the same layer are unconnected. However, such simple neural network structure is difficult to handle sequence data. For example, in natural language processing, it is not enough to comprehend a sentence by understanding its each word. Neural networks are required to process the sequence of these words. The previous input in the sequence will affect the current output, while the network needs to recall the previous information and apply it to the current output calculation. Therefore, the nodes in the same hidden layer are connected, and the input of the hidden layer includes the input layer and the previous hidden layer. Theoretically, RNN can cope with any length sequence data. However, in order to reduce the complexity, the current output is usually related to the current input and the previous several inputs.

Fig. \ref{Fig2} shows a typical RNN structure. The previous input is forwarded to the next hidden layer through the previous hidden layer. When the $n$-bit Ripple Carry Adder is modeled, the carry bit from the previous full-adder will be used as the input of next full-adder. In the VOLtA, the current output is related to the previous and current inputs. Therefore, RNN can model n-RCA and VOLtA with the extremely high modeling accuracy. We will discuss the modeling attacks in detail in Section III.

\begin{enumerate}
  \item [5)]\emph{Evolutionary Strategies (ES)}
\end{enumerate}

ES \cite{Hansen2001,Hansen2016} is a gradient-free stochastic optimization algorithm with invariance under some transformations, parallel scalability and sufficient theoretical analysis. It is appropriate for medium-scale complex optimization problems. ES constantly searches for a normal distribution by iterations. Usually, the normal distribution of iterations is written as $N(m, \sigma^{2}, C)$. $m$ represents the mean of the central position of the distribution; $\sigma$ represents the step size parameter; $C$ represents the covariance matrix. The essence of the ES algorithm is to adjust these three parameters to obtain the best possible search results. How to adjust the step parameters and covariance matrix has a very important impact on the convergence rate of the ES algorithm. The basic idea of adjusting the parameters is to mutate in the direction of the probability of generating a satisfactory solution.
%ES是一种具有良好的不变性(invariance under some transformations)、并行扩展性(scalability)和较为充分的理论分析的无梯度随机优化算法,适用于中等规模的复杂优化问题。进化策略进行黑盒优化的方式是通过反复迭代对一个正态分布进行调整从而不断搜索。其中迭代的正态分布通常情况下写成$N(m, \sigma^{2}, C)$，包含三个参数。其中$m$表示决定分布的中心位置的均值，在算法中决定搜索区域;$\sigma$表示决定分布的整体方差(global variance)的步长参数，在算法中决定搜索范围的大小和强；$C$ 表示决定分布的形状的协方差矩阵，在算法中决定变量之间的依赖关系以及搜索方向之间的相对尺度(scale)。ES 算法的精华所在就是如何调整这三个参数，尤其为了达到尽可能好的搜索效果如何对步长参数和协方差矩阵进行调整，这在ES 算法的收敛速率方面有着相当重要的影响。在通常情况下ES 对参数进行调整的基本思路是调整参数使其朝着产生好解的概率的方向进行变异。
The covariance matrix adaptation evolution strategy (CMA-ES) is a global optimization algorithm developed on the basis of evolution strategy (ES) \cite{Hansen2001}. It combines the reliability and globality of ES with the  adaptiveness of covariance matrices, and can solve complex multiple peak optimization problems. Currently, CMA-ES has attracted much attention in the optimization field due to its exceptional performance and efficient computational \cite{Hansen2016}. In addition, CMA-ES algorithm does not use gradient information in the optimization process. Therefore, as long as the attack model is established, CMA-ES can also effectively attack VOLtA.

\section{Security Analysis and Modeling Attacks on VOLtA }

This section will introduce the VOLtA and analyze its security in detail, and finally the several ML algorithms are proposed to model VOLtA.

\begin{figure}
\centerline{\includegraphics[width=\linewidth]{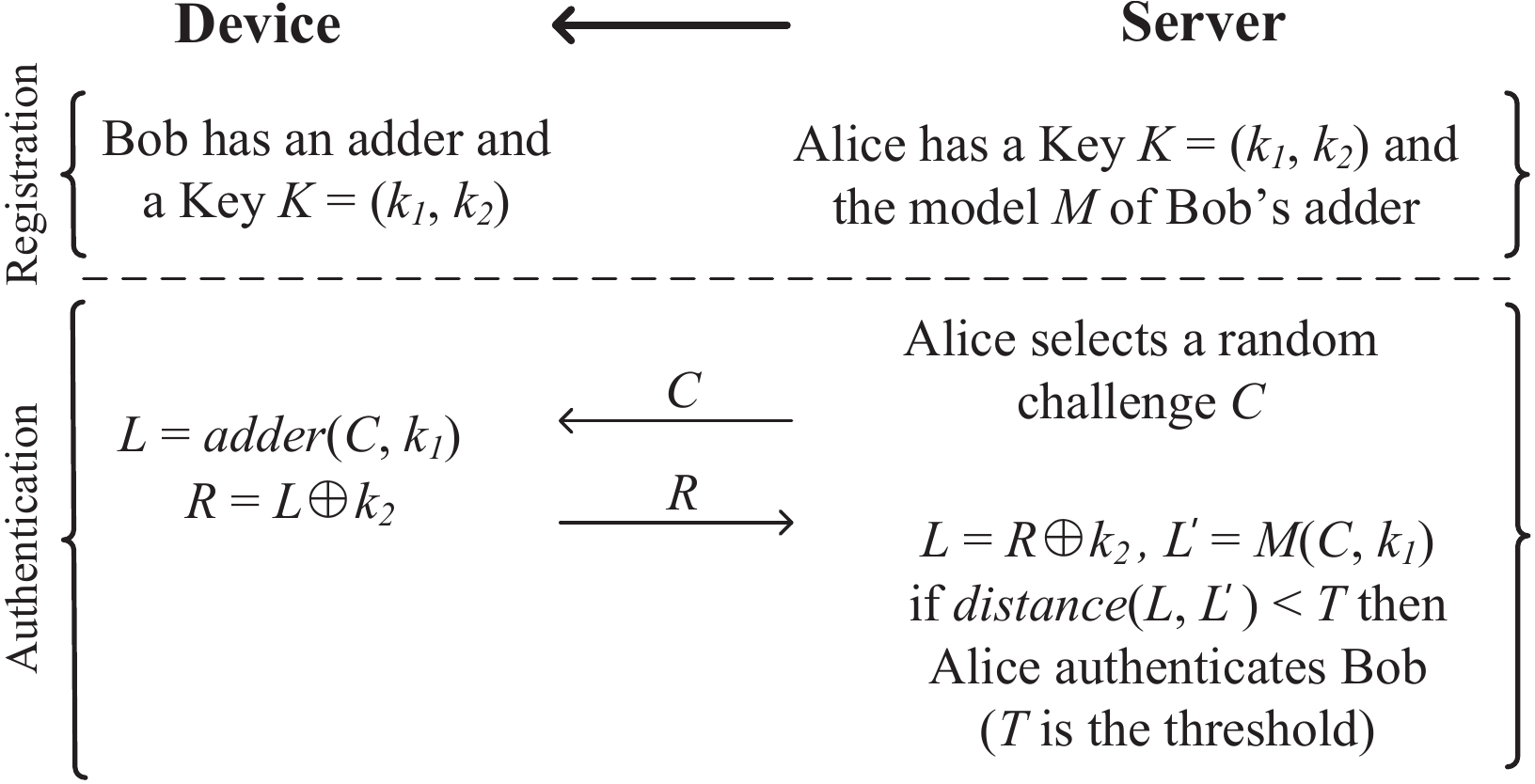}}
\caption{The voltage over-scaling-based lightweight authentication protocol \cite{Arafin2017}, where $adder()$ is the function of adder in VOS and $distance(L,L')$ can be measured by common distance measurement functions such as Hamming distance or Euclidean distance.}
\label{Fig3}
\end{figure}

\subsection{VOLtA}

VOLtA is a two-factor authentication scheme, where two factors include a secret key $K$ and the adder that generates errors in VOS (VOS-adder). The authentication protocol is illustrated in Fig. \ref{Fig3}. Assume that Alice is the server and Bob is the device that carries an adder. The authentication protocol is divided into two phases. In the registration phase, Bob has an adder and a key $K = \{k_1, k_2\}$, Alice has a key $K$ and the adder model $M$ of Bob. In the authentication phase, 1) Alice generates a random challenge $C$ and sends it to Bob; 2) Bob calculates $L = adder (C, k_1)$ using the VOS-adder, then computes $R = L \oplus k_2$, and sends $R$ to Alice; 3) Alice calculates $L = R \oplus k_2$ and $L' = M(C, k_1)$. If the difference between $L$ and $L'$ meets the threshold condition, Alice authenticates Bob.

\subsection{Security Analysis for VOLtA}

In VOLtA, devices must carry the adder and the correct key $K$, otherwise the authentication would be failure. However, the constant key has low obfuscation ability. In addition, the VOS-adder is vulnerable to ML attacks. Therefore, VOLtA suffers the security issues which will be discussed below.

\begin{enumerate}
  \item [1)]\emph{Security Analysis of Constant Key}
\end{enumerate}

As shown in Fig. \ref{Fig1}(a), the inputs of the full-adder are \{$A_1$, $B_1$, $C_{in1}$\}, and the outputs are \{$S_1$, $C_{out1}$\}. Assume that the key $k_1$ is input to $A_1$ and the random challenge $C$ is input to $B_1$. For 1-bit calculation, the input $A_1$ is unchanged because $k_1$ is constant. We can see from Fig. \ref{Fig1}(b), if $A_1 = 0$, then $S_1 = B_1 \oplus C_{in1}$ and $C_{out1} = B_1 \& C_{in1}$; if $A_1 = 1$, then $S_1 =$ $!(B_1 \oplus C_{in1})$ and $C_{out1} = B_1 | C_{in1}$. The full-adder only implements the function of two logic gates after using the constant key $k_1$, which does not increase the difficulty of modeling authentication protocol. We need to model a full-adder without the constant key $k_1$. When the constant key $k_1$ is used, we only need to model the combination of two logic gates. Besides, the VOLtA uses the key $k_2$ to obfuscate the output. In what follows, we will further discuss the obfuscation effectiveness of the key $k_2$.

Assume that $R = L \oplus k_2$, for 1-bit calculation, if $k_2 = 0$, then $R = L$; if $k_2 = 1$, then $R =$ $!L$, which shows that when the output is obfuscated by the constant key, the $i$-th bit output is always unchanged or flipped. For instance, when the adder calculates 4 times, the outputs are $L_{1\sim4} = \{1011_1, 0011_2, ..., 1010_8\}$, the key $k_2 = \{1_1, 0_2, ..., 1_8\}$, and the responses $R_{1\sim4} = \{\underline{0100}_1, 0011_2, ..., \underline{0101}_8\}$ after using the XOR obfuscation. Obviously, when the $i$-th bit key $k_{2, i} = 1$, the $i$-th bit response is inverted such as the underlined parts of $R_{1\sim4}$; when the $i$-th bit key $k_{2, i} = 0$, the $i$-th bit response remains unchanged. We just need to establish a ML model for the $i$-th bit output to implement similar functions.

As analyzed above, the defenses that use constant keys to obfuscate the output is unable to resist ML attacks.

\begin{enumerate}
  \item [2)]\emph{Complexity of Challenge-response Mapping}
\end{enumerate}

\begin{figure}
\centerline{\includegraphics[width=\linewidth]{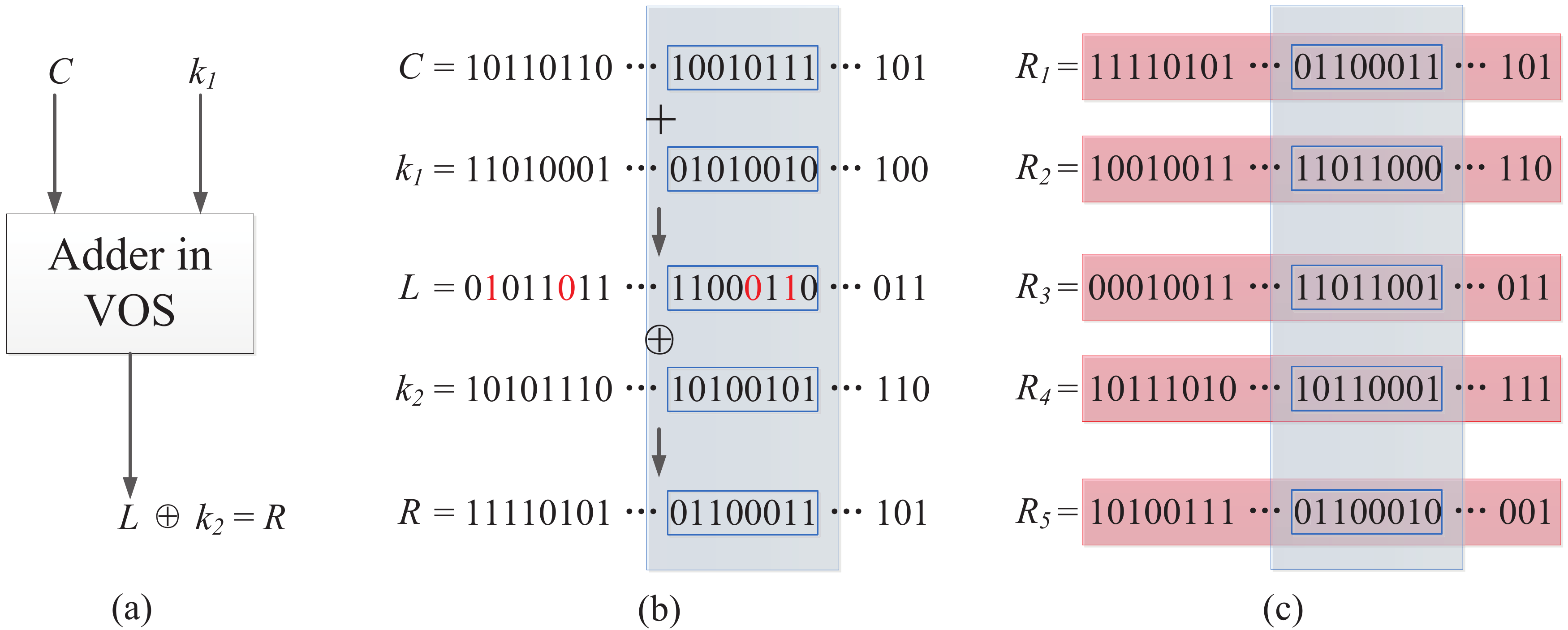}}
\caption{A calculation example of VOLtA. In (a), the challenge $C$ and the key $k_1$ are calculated by the VOS-adder to generate $L$, then $L$ and the key $k_2$ are XORed to generate response $R$. An example of the computing process is given in (b), in which the red numbers indicate the computing errors. The response example of 5 times authentication is shown in (c). We call the data in red box as horizontal data, and the data in blue box as vertical data.}
\label{Fig4}
\end{figure}

\begin{figure*}
\centerline{\includegraphics[width=\linewidth]{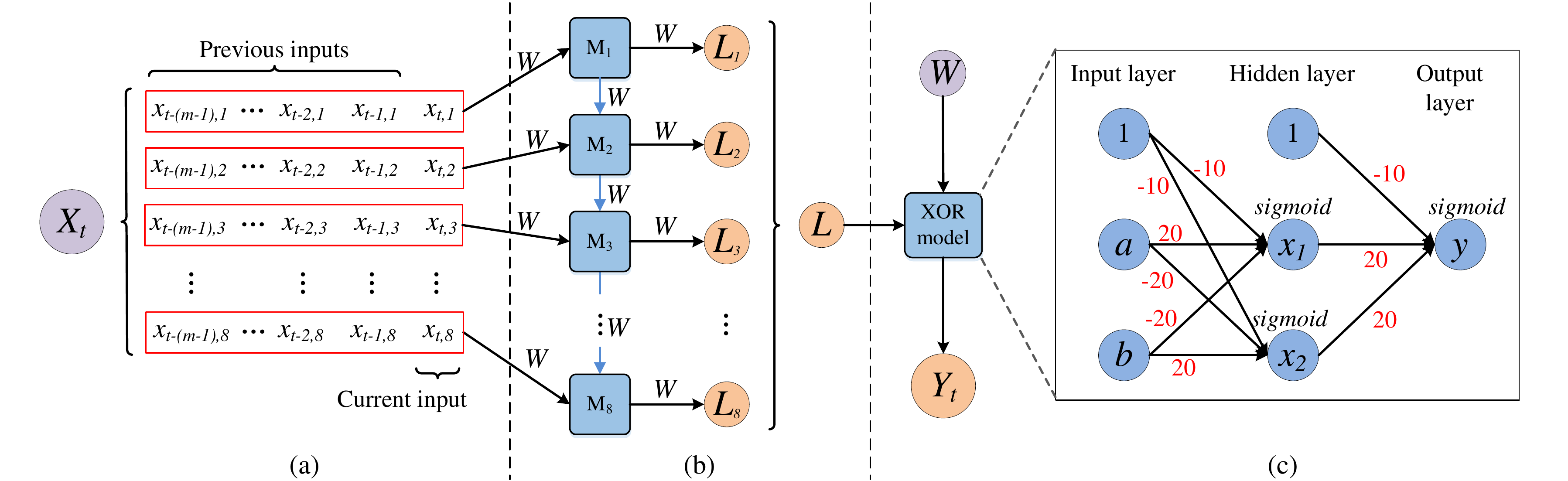}}
\caption{The attack model of VOLtA.}
\label{Fig5}
\end{figure*}

The VOLtA employs the CRPs to authenticate devices. The mapping of challenge-response (CR) depends on the calculation errors generated by a VOS-adder. As long as the effective and enough CRPs are collected, ML algorithms can model the VOS-adder to simulate its CR behavior. In the VOLtA \cite{Arafin2017}, assume that the length of the random challenge $C$ is 8*$n$ bits, the $K$ is 16*$n$ bits ($k_1$ and $k_2$ are both 8*$n$ bits), which incurs unacceptable key storage overhead. For example, if a 52*40 pixels image is used as the challenge for authentication, the required key $K$ will be 16*52*40 = 33,280 bits. In the case of ignoring the key storage overhead, we discuss the complexity of CR mapping.

A calculation example of VOLtA is illustrated in Fig. \ref{Fig4}. The adder performs each addition and XOR operation with the corresponding $k_1$ and $k_2$. Therefore, the horizontal data are obfuscated by different keys so that horizontal data cannot be used to train the model with high accuracy. However, from the perspective of vertical data, the key used by the $i$-th byte of $C$ is the same for each time, and the calculation of the data in the blue box (see Fig. \ref{Fig4}(b) and Fig. \ref{Fig4}(c)) uses the same key. Therefore, we can use the data in the blue box to model the operation of its corresponding byte, and the VOLtA can be modeled using valid CRPs with high prediction accuracy.

%Therefore, in the two-factor authentication scheme, both the key and the errors produced by the VOS-adder can be modeled by ML, i.e., the VOLtA can be modeled successfully by using valid CRPs. In what follows, we will discuss the ML modeling attacks in detail.

\subsection{Modeling Attacks on VOLtA }

As analyzed above, we need to model the logic gates first. The common logic gates include NOT gate, AND gate, OR gate and XOR gate, where the XOR gate is linearly inseparable and hence it is often used to encrypt information in cryptography. However, the XOR can be implemented by other logic operations. For example,

\begin{equation}
a \oplus b = (a \& !b) | (!a \& b)
\end{equation}

\noindent where '$!$' is NOT, '$\&$' is AND, '$|$' is OR and '$\oplus$' is XOR. Besides, NOT, AND, OR and XOR can be approximated as:

\begin{equation}
!a = 1 - a
\end{equation}
\begin{equation}
a \& b \approx f_{and}(a,b) = sigmoid(20*a + 20*b - 30)
\end{equation}
\begin{equation}
a | b \approx f_{or}(a,b) = sigmoid(20*a + 20*b - 10)
\end{equation}
\begin{equation}
a \oplus b \approx f_{xor}(a,b) = f_{or}(f_{and}(a,1-b),f_{and}(1-a,b))
\end{equation}

\noindent where $sigmoid(x)=1/(1+ e^{-x})$, which is a common activation function in the neural network. Substituting Eqn. (4), (5) and (6) into Eqn. (3), the approximate Eqn. (7) for XOR can be obtained. Based on this, we design the neural network structure shown in Fig. \ref{Fig5}(c) to model the XOR gate, where $x_1 \approx a \& !b$, $x_2 \approx !a \& b$ and $y \approx x_1 | x_2$. To model the required functions, we expand the number of neurons in the hidden layer to 10, and set the edges with random weight parameters to model any logic gate; when the obfuscation mechanism which employs the constant key is modeled, the weight of edges is set to the red numbers in Fig. \ref{Fig5}(c) and neuron $b$ is set to a random parameter.

%Although the VOLtA authenticates devices with errors produced by the VOS-adder, the error rate is low. Thus, we expand the neuron number of the normal adder model to model the VOS-adder. A $n$-bit Ripple Carry Adder ($n$-RCA) is shown Fig. \ref{Fig6}, in which the input $C_i$ of FA$_i$ is also the output of FA$_{i-1}$, which is the same as RNN structure. Therefore, the $n$-RCA model is the combination of $n$ head-to-tail full-adder models.

The attack model of VOLtA is shown in Fig.\ref{Fig5}. Since the current output in VOLtA is related to the current input and the previous input, the input of the model is adjusted to learn the effective mapping between input and output. As shown in Fig. \ref{Fig5}(a), the current input is combined with the previous input to create the actual input $X_t = \{x_{t-(m-1)}, ..., x_{t-2}, x_{t-1}, x_t\}$, where $m$ denotes the number of input bytes, $x_t$ denotes $t$ timing input, and $x_{t-m, i}$ denotes the $i$-th bit of $t-m$ timing input. We use the vertical data to model the VOLtA. The neural network model of 8-bit Ripple Carry Adder (8-RCA) is shown in Fig. \ref{Fig5}(b), the $i$-th bit of 8-RCA is input to $M_i$, and $M_{i-1}$ serves as the input of $M_i$, which is a typical RNN structure. The XOR obfuscation mechanism is shown in Fig. \ref{Fig5}(c). All weight parameters $W$ are random numbers that need to be adjusted.

\section{Challenge Self-obfuscation Structure}

To resist ML attacks, this paper proposes a \underline{c}hallenge \underline{s}elf-\underline{o}bfuscation \underline{s}tructure (CSoS) against ML attacks. This section will introduce the CSoS and the CSoS-based authentication protocol for VOLtA in detail. In addition, the hardware implementation and security analysis of CSoS for VOLtA and Arbiter PUF will be introduced.

\subsection{The CSoS}

\begin{figure}
\centerline{\includegraphics[width=\linewidth]{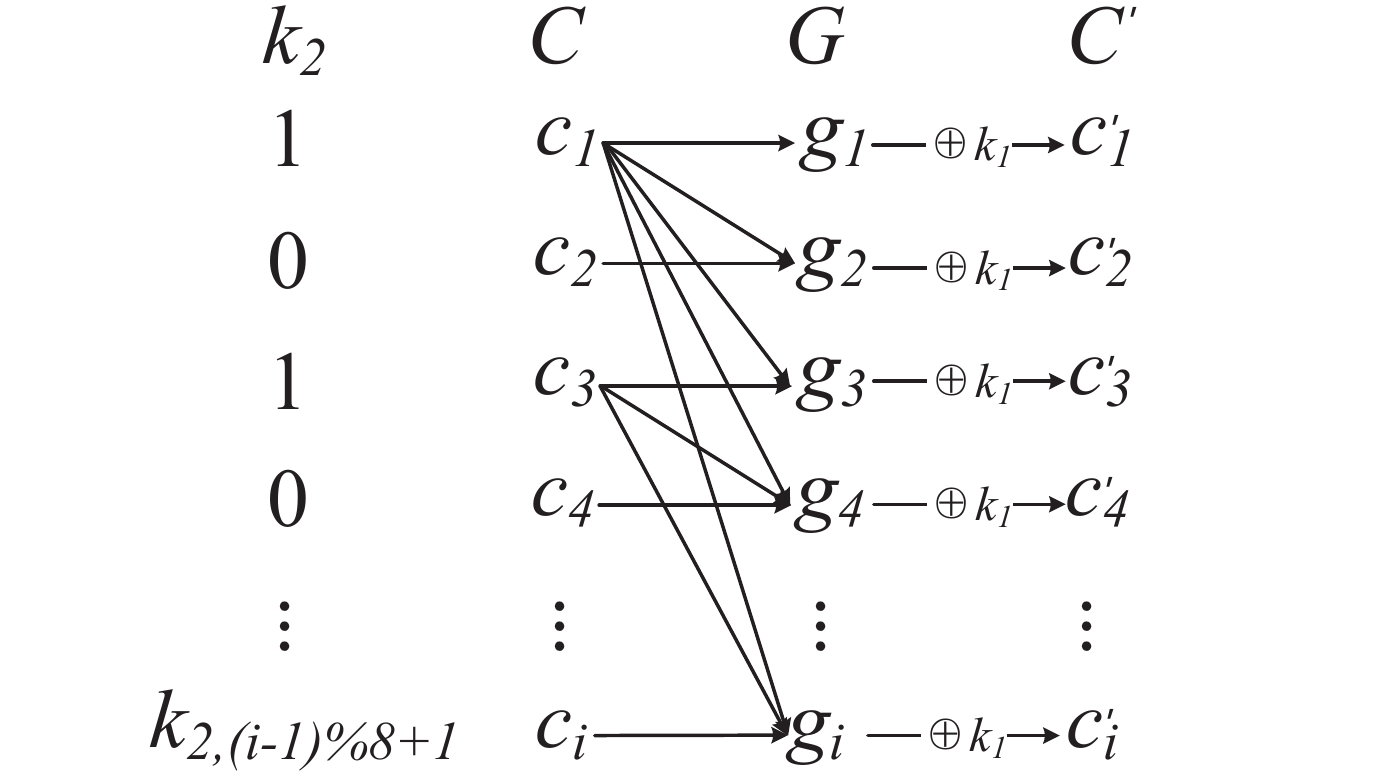}}
\caption{The obfuscation process of CSoS}
\label{Fig8}
\end{figure}

The errors generated by the VOS-adder are related to input timing, and the current output is determined by the current input and the previous input. If the correlation among inputs is enhanced or the input is obfuscated, ML modeling attacks would be difficult.

The key idea of CSoS is to combine the previous input with secret keys and random numbers to generate dynamic new keys, and exploit the new keys to obfuscate the current input. The 8-RCA is used as an example, assume that the challenge $C = \{c_1, c_2, ..., c_t\}$, the keys are $k_1$ and $k_2$, and the obfuscated challenge $C'=\{c_1',c_2', ...,c_t'\}$ can be expressed as:

\begin{equation}
c_i'=k_1 \oplus g_i
\end{equation}
\begin{equation}
g_i =f(c_1,k_{2,1}) \oplus f(c_2,k_{2,2})\oplus ... \oplus f(c_{i-1},k_{2,(i-2)\%8+1}) \oplus c_i
\end{equation}
\begin{equation}
f(x, y)  =
\left\{
             \begin{array}{lr}
             x, & \mbox{if}\ y = 1  \\
             00...00, & \mbox{if}\ y = 0
             \end{array}
\right.
\end{equation}

In Eqn. (9), an 8-bit key $k_2$ is used to obfuscate the intermediate calculation values $G = \{g_1, g_2, ..., g_t\}$. For instance, if $k_2 = 10100101$ and $k_{2, i}$ denotes the $i$-th bit of $k_2$. The obfuscation process of CSoS is shown in Fig. \ref{Fig8}, where the connection between $c_i$ and $g_i$ indicates XOR, i.e., $g_3$ is connected with \{$c_1, c_3$\} to indicate that $g_3 = c_1 \oplus c_3$ and $g_4$ is connected with \{$c_1, c_3, c_4$\} to indicate that $g_4 = c_1 \oplus c_3 \oplus c_4$. Since the attackers do not know the $k_1$ and $k_2$, it is impossible to collect the relevant information of obfuscated challenge $C'$. In the authentication, the obfuscated challenge $C'$ will be transmitted as the real challenge to the adder for calculation. Attackers can only collect the challenge $C$ and the response corresponding to $C'$. In this case, the attackers cannot collect valid CRPs for modeling attacks.

\subsection{The CSoS-based Authentication Protocol}

\begin{figure}
\centerline{\includegraphics[width=\linewidth]{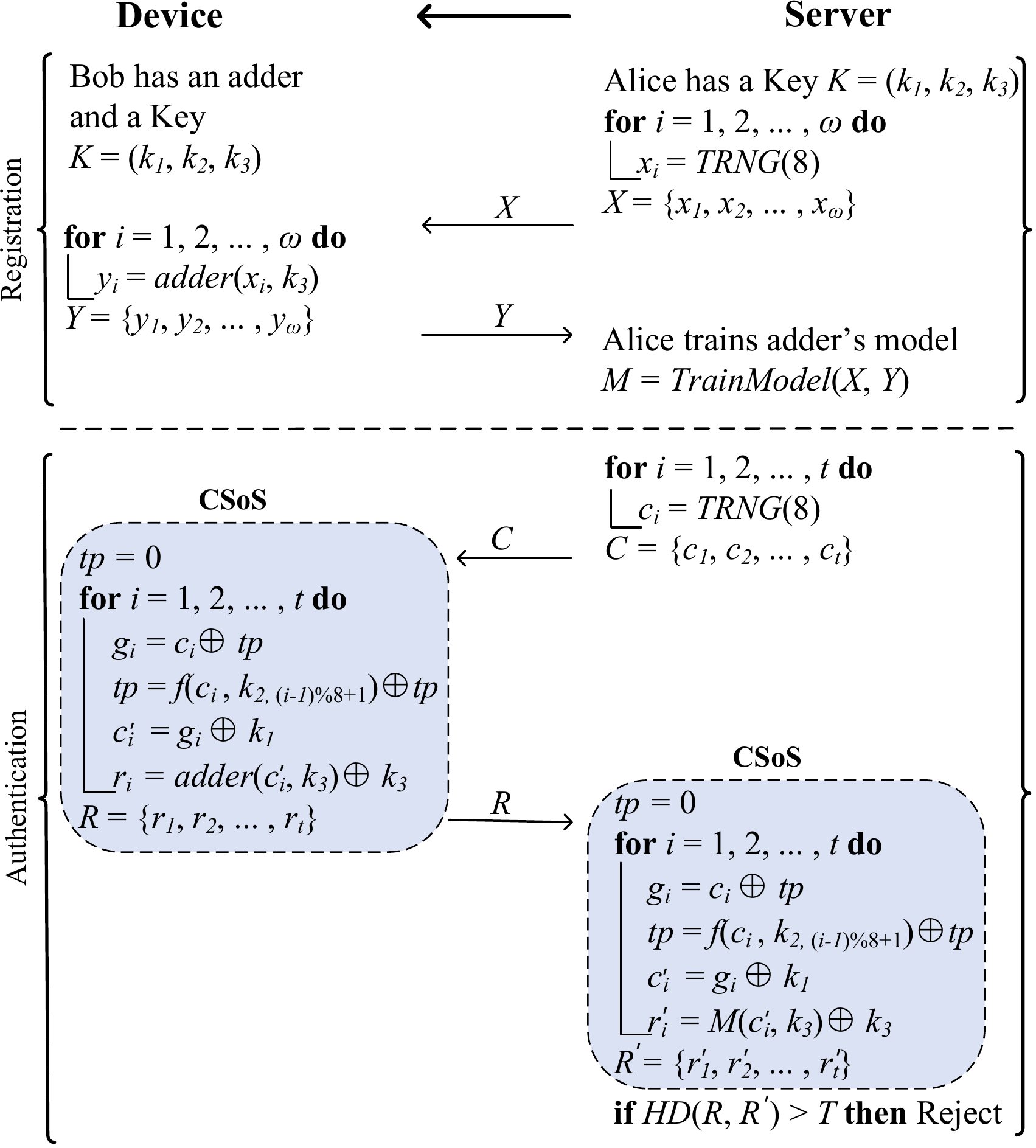}}
\caption{The CSoS-based ML attacks resistant authentication protocol.}
\label{Fig10}
\end{figure}

We propose a CSoS-based ML attacks resistant authentication protocol for VOLtA. The key $K$ and the VOS-adder are used to authenticate devices. The key $K$ consists of three different keys $k_1$, $k_2$ and $k_3$, where $k_1$ and $k_2$ are used to obfuscate the challenge in CSoS, and $k_3$ has two functions: 1) used as an input of the adder; 2) encrypting the output of adder with the XOR operation. The length of $k_1$ and $k_3$ are 8 bits, and $k_2$ can be any length (in this paper, $k_2$ is set to 8 bits). As shown in Fig. \ref{Fig10}, the authentication protocol includes registration and authentication:

\begin{flushleft}
\textbf{Registration}
\end{flushleft}

\begin{itemize}

\item [i.]
Alice and Bob obtain the secret key $K = \{k_1, k_2, k_3\}$ through key sharing or other similar methods;
\item [ii.]
Alice randomly generates an input bitstream $X = \{x_1, x_2, ..., x_{\omega}\}$, where $\omega$ is the number of bytes of $X$, then sends $X$ to Bob;
\item [iii.]
Bob adds $x_i$ and $k_3$ using VOS-adder to generate an output bitstream $Y = \{y_1, y_2, ..., y_{\omega}\}$, and sends $Y$ to Alice;
\item [iv.]
Alice uses $X$ and $Y$ to train the adder model of Bob.
\end{itemize}

\begin{flushleft}
\textbf{Authentication}
\end{flushleft}

\begin{itemize}

\item [i.]
Alice generates a random challenge $C = \{c_1, c_2, ..., c_t\}$, and sends it to Bob;
\item [ii.]
Bob employs CSoS to obfuscate challenge $C$ to get the challenge $C'=\{c_1',c_2', ...,c_t'\}$, and adds $c_i'$ and $k_3$  using VOS-adder, then XORs the calculation result and $k_3$ to obtain the response $R = \{r_1, r_2, ..., r_t\}$, and finally $R$ is sent to Alice;
\item [iii.]
Alice obtains the obfuscated challenge $C'$ through CSoS and $C$, then employs the model $M$ and $k_3$ to generate the response $R'$;
\item [iv.]
Alice calculates the Hamming distance $HD(R, R')$ between $R$ and $R'$. If the $HD(R, R')$ is greater than the threshold condition, the authentication fails.
\end{itemize}

\subsection{Hardware Implementation}
%
%\begin{figure}[h]
%\centerline{\scalebox{0.8}{\includegraphics[width=\linewidth]{Fig9}}}
%\caption{The hardware implementation of CSoS.}
%\label{Fig9}
%\end{figure}

\begin{figure}
\centerline{{\includegraphics[width=\linewidth]{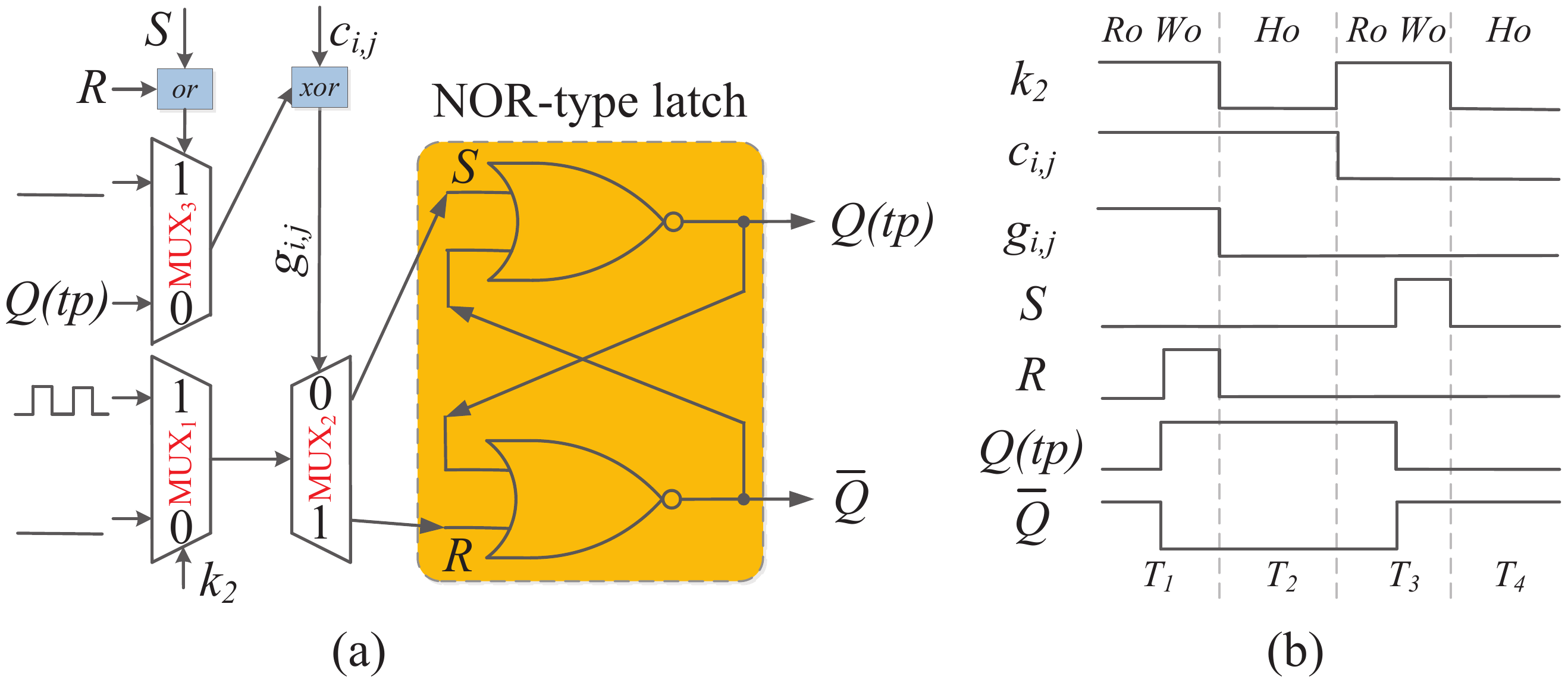}}}
\caption{(a) The 1-bit input cache structure (ICS). (b) An example of ICS.}
\label{ICS}
\end{figure}

In Eqn. 8, $g_i$ need to be stored temporarily in the calculation. Therefore, we design the input cache structure (ICS), as shown in Fig. \ref{ICS}(a), which consists of some latches and multiplexers (MUXs). A NOR-type latch is used to store 1-bit $g_i$ and the truth table is given in Table I. When $S = R = 0$, the circuit remains in its original state; when $S = 0, R = 1$, regardless of the state of $Q$ and $\overline{Q}$, there will be $Q = 1 , \overline{Q} = 0$; when $S = 1, R = 0$, regardless of the state of $Q $ and $ \overline{Q}$ , there will be $Q = 0 , \overline{Q} = 1 $. It is worth noting that $S = R = 1$ cannot be employed as an input signal.

\begin{table}
\caption{The truth table for NOR-type latch}
\centering
\begin{tabular}{|c|c|c|c|c|c|}
\hline
$S$ & $R$ & $Q$ & $\overline{Q}$ & $Q^{'}$ & $Functiong$\\
\hline
$0$ & $0$ & $0$ & $1$ & $0$ & Hold\\
\hline
$0$ & $0$ & $1$ & $0$ & $1$ & Hold\\
\hline
$0$ & $1$ & $0$ & $1$ & $1$ & Set to 1\\
\hline
$0$ & $1$ & $1$ & $0$ & $1$ & Set to 1\\
\hline
$1$ & $0$ & $0$ & $1$ & $0$ & Set to 0\\
\hline
$1$ & $0$ & $1$ & $0$ & $0$ & Set to 0\\
\hline
$1$ & $1$ & $0$ & $1$ & $-$ & $-$\\
\hline
$1$ & $1$ & $1$ & $0$ & $-$ & $-$\\
\hline
\end{tabular}
\end{table}

1-bit input cache structure (ICS) is shown in Fig. \ref{ICS}(a). we take the $j$-th bit of $g_i$ as an example $g_{i,j}$, the ICS includes three operations:

\begin{itemize}
\item \textbf{Read operation (Ro):}
The NOR-type latch keeps latching state and outputs $tp$ before calculating $g_{i,j}$, so that $g_{i,j} = c_{i,j} \oplus tp$.
\item \textbf{Write operation (Wo):}
After calculating $g_{i,j}$, the NOR-type latch is released from the latching state and then $g_{i,j}$ is written into the NOR-type latch, i.e., $tp = g_{i,j}$.
\item \textbf{Hold operation (Ho):}
The NOR-type latch holds latching state and outputs $tp$ throughout, $g_{i,j} = c_{i,j} \oplus tp$, and $tp$ keeps unchanged until the next operation is performed.
\end{itemize}

The read, write and hold operations are controlled by a signal based on the key $k_2$. We assume $k_2 = 10100101$, and the control signals of ICS are \underline{10100101} \underline{10100101} ... \underline{10100101}. If the control signal is '1', ICS performs the read and write operation; if the control signal is '0', ICS executes the hold operation. We use the NOR-type latch combined with three MUXs to implement these operations. Fig. \ref{ICS}(b) gives a instance of storing $g_i$. Assuming that the single signal duration of $k_2$ is $T$, the '1' port of $MUX_1$ is a periodic signal $ps$ with a period of $T$, and '0' port is a low level signal. In the first half of time $T_1$, $k_2 = 1, c_{i,j} = 1$, $ps$ is connected to the circuit and transferred to $R$; in the second half of time $T_1$, $S = 0, R = 1$, there is $Q(tp) = g_{i,j} = 1$. In addition, when executing a single write operation, we use $MUX_3$ to prevent the updated value of $tp$ from affecting the value of $g_i$ again. Similarly, in the time period $T_3$, $g_{i, j} = 0$ is updated to $Q(tp)$. In this way, we get $G = \{g_1, g_2, ..., g_t\}$ which is obfuscated by the key $k_2$. In the obfuscation process, the CSoS just combines the previous input with keys to obfuscate the current input, and hence does not affect the original uniqueness and reliability of circuit.

%我们在NOR-type latch的基础上加了三个MUXs实现整个操作，Fig. \ref{ICS}(b) shows $g_i$的储存实例，假设$k_2$的单个信号时长为$T$，$MUX_1$'1'端是一个周期为$T$的周期信号$ps$，'0'端为低电平信号。在时间$T_1$的前半段，$k_2 = 1, c_{i,j} = 1$时，$ps$ 连入电路并且传输到$R$；在时间$T_1$ 的后半段，$S = 0, R = 1$，从而有$Q(tp) = g_{i,j} = 1$. 除此之外，在执行单个Wo时，我们使用$MUX_3$ 来防止$tp$的值更新后再次对$g_i$ 的值造成影响。同理在时间段$T_3$，$g_{i,j} = 0$ 被更新到$Q(tp)$。In this way, we get $G = \{g_1, g_2, ..., g_n\}$ which is obfuscated by the key $k_2$.

\begin{figure}
\centerline{{\includegraphics[width=\linewidth]{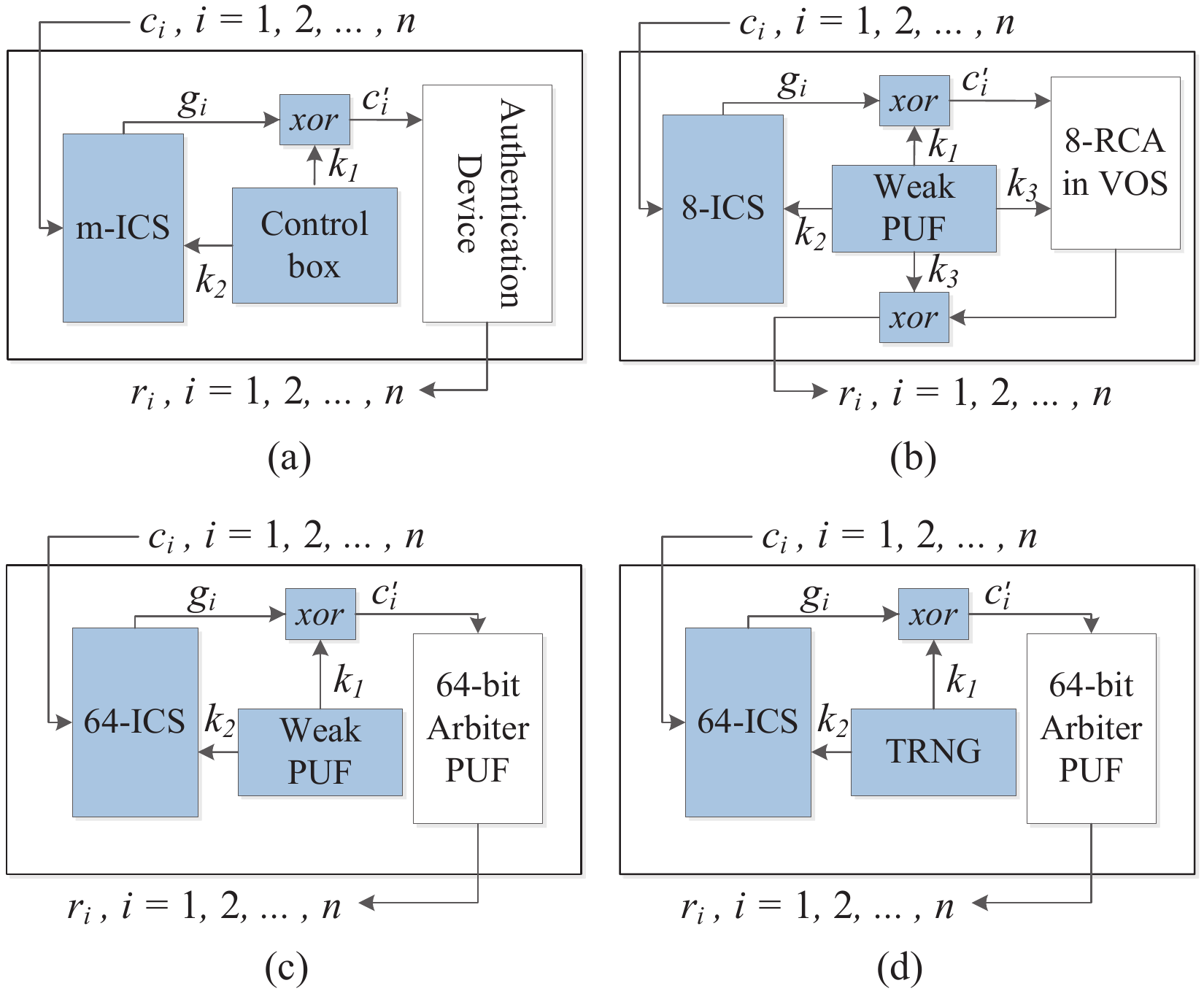}}}
\caption{(a) The hardware implementation of CSoS. (b) The CSoS for an 8-RCA in VOS. (c) The WCSoS for a 64-bit Arbiter PUF. (d) The TCSoS for a 64-bit Arbiter PUF.}
\label{Fig9}
\end{figure}

As shown in Fig. \ref{Fig9}(a), the CSoS proposed in this paper consists of the ICS, the control box and some XOR gates. The key generator is used to generate the key $k_1$ and $k_2$ for obfuscation. It can be implemented using Weak PUF and True Random Number Generator (TRNG), named Weak PUF-based CSoS (WCSoS) and TRNG-based CSoS (TCSoS) respectively. Fig. \ref{Fig9}(b) gives the deployment of CSoS in VOLtA, which is corresponding to Section IV.B. It is worth noting that the CSoS is a universal obfuscation method and hence can also be used for Strong PUFs. In Fig. \ref{Fig9}(c) and \ref{Fig9}(d), a classic Strong PUF, 64-bit Arbiter PUF, is used as an example to deploy WCSoS and TCSoS. In the WCSoS, $k_1$ and $k_2$ are different keys generated by Weak PUF. In the TCSoS, $k_1$ and $k_2$ are random numbers generated by TRNG. In order to reduce the complexity of authentication, we make $k_1$ equal to $k_2$. Moreover, the TCSoS does not require the key storage and has higher security. For example, if the number of bits in the TRNG is $T_{num} = 4$ and TRNG(4) = 1010, $k_1 = k_2 = \{\underline{1010}\ \underline{1010}\ ...\ \underline{1010}\}$ has a total of 64 bits. In authentication, 64$\times$64-bit challenges are input to the device in the time series to generate 64-bit responses which are sent to the server. Then the server needs to enumerate all the possibilities of $k_1$ and $k_2$ and verifies these responses one by one to authenticate the device (the number of possibilities is $2^{T_{num}}$).

%The control box最要用于生成控制信号$k_2$以及用于异或混淆的密钥$k_1$，可以使用Weak PUF和True Random Number Generator (TRNG)实现，分别叫做Weak PUF-based CSoS (WCSoS) 和TRNG-based CSoS (TCSoS)。 我们在VOLtA 中部署了weak PUF-based CSoS，如Fig. \ref{Fig9}(b) 所示，它的混淆机制对应本章节的第二部分。值得注意的是，CSoS具有较好的通用性，它几乎适用于所有的String PUF。在Fig. \ref{Fig9}(c)(d)中，我们以经典的String PUF(64-bit Arbiter PUF) 为例，对其部署了WCSoS和TCSoS。TCSoS相较于WCSoS 的主要区别在于$k_1$ 和$k_2$ 的值，在WCSoS 中，$k_1$ 和$k_2$的值是由Weak PUF生成的不同密钥；而在TCSoS 中，$k_1$ 和$k_2$的值是由TRNG 生成的随机数，并且在我们结构中为了减少认证的复杂度，令$k_1 = k_2$。假设TRNG的位数为$T_{num} = 4$, 在一次认证中TRNG(4) = 1010, 那么$k_1 = k_2 = \{\underline{1010}\ \underline{1010}\ ...\ \underline{1010}\}$共64bit。 在认证时，激励为64个64位的二进制串按时序输入，对应输出64bit 响应。

 %Second, the circuit XORs $g_i$ and $k_1$ to get $C'=\{c_1',c_2', ...,c_t'\}$.  Finally, the circuit adds $c_i'$ and $k_3$ using 8-RCA in VOS, and XORs the calculation result and $k_3$ to obtain response $R = \{r_1, r_2, ..., r_n\}$.

\subsection{Security Analysis}
In this study, we assume that the server is trustworthy and the attacker cannot get the keys and the cloned model stored in the server.
\subsubsection{Key Security}
Our proposed CSoS combines the previous input with the secret keys or random numbers to obfuscate the current input. In the TRNG-based CSoS for Arbiter PUF, if attackers know the cloned model of Arbiter PUF, they can enumerate all the random numbers to clone the authentication protocol. However, the cloned model of Arbiter PUF is securely stored in the server and hence will not be leaked. In the weak PUF-based CSoS, key generator on the device can be implemented with the weak PUF \cite{Zhang2014,Pang2017,Ma2018}. If attackers get the secret keys, the authentication protocol would be broken. Side-channel attacks are powerful noninvasive attacks that exploit the leakage of physical information when the encryption algorithm is being executed on a system \cite{Kocher1996}. Several side-channel attacks on weak PUFs have been reported within the past couple of years \cite{Merli2011a,Merli2013}, and most of the authors have pointed out potential countermeasures to their proposed attacks. We don't propose any solution to prevent side-channel attacks on weak PUFs because it is beyond the scope of this article.

%In our proposed obfuscation structure,  or the TRNG \cite{Zhang2018}, which can improve the security of keys. the attackers from stealing the keys from the device to restore the obfuscated challenge to perform ML modeling attacks.

\subsubsection{Brute Force Attacks}
Attackers enumerate the keys and build multiple models to attack. In the weak PUF-based CSoS for VOLtA, assume that the keys $k_1$ and $k_3$ are 8 bits, $k_2$ is $x$ bits, the number of models that attackers need to build to pass the authentication is $2^{(16+x)}$ which is increased exponentially with the increasing of $x$. In the TRNG-based CSoS, the CSoS uses the TRNG to generate keys $k_1$ and $k_2$ ($k_1=k_2$) to improve security, which only increases the computational overhead of server in authentication. In this case, the number of models that the attackers need to establish is related to the number of collected CRPs. The attackers need to select effective training set in massive data and build an efficient model. Therefore, it is impossible for attackers to clone the CSoS-based authentication by brute-force attacks.

\subsubsection{Learning-based Attacks}
Attackers try to collect large amounts of data to conduct ML attacks. The function of Arbiter PUF can be represented by an additive linear delay model, and the mathematical model of the Arbiter PUF is described in \cite{Rührmair2010,Lim2005}. In this model, we can define the final delay difference $\Delta$ between the upper and the lower path (see Fig. \ref{APUF}) as:

\begin{equation}
\Delta = \Omega \cdot \Phi(C)
\end{equation}
where $\Omega = \{ {\omega^1,\omega^2,...,\omega^n,\omega^{n+1}}\}$, the dimensions of $\Omega$ and $\Phi$ are both $n + 1$. The parameter vector $\Omega$ represents the delay of each stage in an Arbiter PUF; the eigenvector $\Phi(C) = (\phi^1(c),...,\phi^n(c),1)^T$ represents a function with the $n$-bit challenge, while $\phi^l(\cdot)$ is a function that can be represented by

\begin{equation}
\phi^l(c) = \prod_{j = l}^n (1-2c_j), l = 1,..., n
\end{equation}

The vector $\Omega$ determines a separate hyperplane in all the eigenvectors by $\Omega \cdot \Phi(C) = 0$. Any challenges have their vectors $\Phi(C)$ located
on one side of the hyperplane produce $\Delta < 0$, and on the other side produce $\Delta > 0$. Note that there is non-linear relationship between the challenge $C = (c_1, c_2, ..., c_n)$ and delay difference $\Delta$, but the feature vector $\Phi(C) = (\phi^1(c),...,\phi^n(c),1)$ is linearly related to $\Delta$. This makes the application of ML very effective \cite{Rührmair2010,Zhang2018,Ye2017}.

However, in the CSoS-based Arbiter PUF, the $i$-th timing challenge $C_i' = (c_{i,1}', c_{i,2}', ..., c_{i,n}')$, and the final delay difference $\Delta$ can be represented as:

\begin{equation}
\Delta = \Omega \cdot \Phi(C_i')
\end{equation}
where $\Phi(C_i') = (\phi^1(c_i'),...,\phi^n(c_i'),1)$ is a feature vector, and

\begin{equation}
\phi^l(c_i')=\prod_{j = l}^n (1-2c_{i,j}'),l = 1,..., n
\end{equation}
according to Eqn. (8),(9) and (10),

\begin{equation}
\begin{split}
c_{i,j}' &=k_{1,j} \oplus f(c_{1,j},k_{2,1}) \oplus f(c_{2,j},k_{2,2})\oplus ...  \\
         &\ \ \  \oplus f(c_{i-1,j},k_{2,i-1}) \oplus c_{i,j}\\
         &=Prefix_{i,j} \oplus c_{i,j}
\end{split}
\end{equation}
where $c_{i,j}$ represents the $i$-th timing and $j$-th bit of challenge. $x \oplus y$ can be expressed by Eqn. (16)

\begin{equation}
x \oplus y = x + y - 2 x \cdot y
\end{equation}
Therefore, the Eqn. (14) for CSoS can be represented as

\begin{equation}
\begin{split}
\phi^l(c_i') &=\prod_{j = l}^n (1-2c_{i,j}')\\
             &=\prod_{j = l}^n (1-2( Prefix_{i,j} + c_{i,j} - 2Prefix_{i,j} \cdot c_{i,j} ))\\
             &=\prod_{j = l}^n (1-2c_{i,j})(1-2Prefix_{i,j})\\
             &=\prod_{j = l}^n (1-2c_{i,j}) \cdot \prod_{j = l}^n (1-2Prefix_{i,j})
\end{split}
\end{equation}

We can see from Eqn. (17), the challenges in $i$-th timing are obfuscated by keys and previous challenges ($Prefix_{i,j}$) in the CSoS. Even if the challenges are same, the generated obfuscated challenges may be different due to the different previous challenges. Furthermore, some previous challenges are hidden by keys and not used to obfuscate the current challenge. In our experiments, RNN fails to attack the CSoS without knowing which previous challenges are used. Therefore, it is difficult for attackers to model it with ML methods due to the high complexity of the obfuscated CRP mapping.

\section{Experiments and Results}

\subsection{Experimental Setup and Data Collection}

We have reproduced the simulation experiments for a 8-RCA circuit in \cite{Arafin2017} and performed simulations in the HSpice platform using the FreePDK 45nm libraries \cite{Stine2007}. The python 3.6.4 programming language and the tensorflow 1.6.0 neural network toolkit are used to conduct modeling attacks. All experiments are conducted on the Intel(R) Core(TM) i5-7400 CPU @ 3.00GHz, 8G RAM and GeForce GT 720 GPU.

We modify the threshold voltages of the NMOS and PMOS models in the FreePDK 45nm libraries to simulate process variations based on the Gaussian Distribution $\pm$7$\%$. The circuit netlist for the 8-RCA is designed by using the modified NMOS and PMOS models at random, and then the circuit simulation is implemented in HSpice, where the simulation temperature is 25$^{\circ}$C. We collect the challenge-response pairs (CRPs) generated randomly by this 8-RCA to perform modeling attacks. In addition, we get the first 18 bytes of random challenge $C$ as vertical data (for the definition of vertical data, see Section III). To get massive vertical data more efficiently, the vertical data is arranged as a bit-stream for collection. In this bit-stream, the signal LOW is maintained for a period time after completing the 18 bytes calculation, and then the 18 bytes computing is performed again, while massive vertical data can be produced by such a loop.

%However, due to the limited memory and computing power of computer, Hspice only simulated 20,000 pairs of CRPs, which are not enough to verify the effectiveness of CSoS. Therefore, we also carried out simulation experiments on WCSoS and TCSoS Arbiter PUF. Unlike VOLtA, in our simulation, the delay of the multiplexer segment of Arbiter PUF is generated by Gaussian Distribution, which follows the well-established linear additive delay model for PUFs \cite{Lee2004,Lim2005}. In addition, we simulate the TRNG function with the $random.randint()$ function in Python. $10^6$ CRPs are simulated in the Arbiter PUF experiments.
 We use Hspice to simulate 20,000 CRPs for CSoS. We also carry out simulation experiments on WCSoS and TCSoS Arbiter PUF. In our simulation, the delay of the multiplexer segment of Arbiter PUF is generated by Gaussian Distribution, which follows the well-established linear additive delay model for PUFs \cite{Lee2004,Lim2005}. In addition, we simulate the TRNG function with the $random.randint()$ function in Python. $10^6$ CRPs are simulated in the Arbiter PUF experiments.

%由于设备内存和运算能力有限，Hspice仅仿真了20000 对CRPs，并且20000对CRPs还远不足以验证CSoS的抗机器学习效应，因此我们对WCSoS 和 TCSoS Arbiter PUF 也进行了仿真实验。不同于VOLtA 的是，在仿真时，Arbiter PUF 中多路复用器中路段的延迟我们用高斯分布模拟产生，此方法遵循成熟的累加的线性延迟模型(well-established linear additive delay model for PUFs)\cite{Lee2004,Lim2005}。 另外，我们用Python 中的random.randint() 函数模拟TRNG 功能。$10^6$ 对CRPs 被收集用于进行Arbiter PUF的实验。

\subsection{Attacks}

ANN, RNN and CMA-ES are used to evaluate the effectiveness of modeling attacks, and RNN is used to attack the VOLtA and the no-key-VOLtA (VOLtA without keys). 20,000 CRPs for VOLtA and no-key-VOLtA are simulated by using HSpice. ML models are trained by using 10,000 CRPs and the rest of 10,000 CRPs are used as the testing set.

\begin{figure}
\centerline{\includegraphics[width=\linewidth]{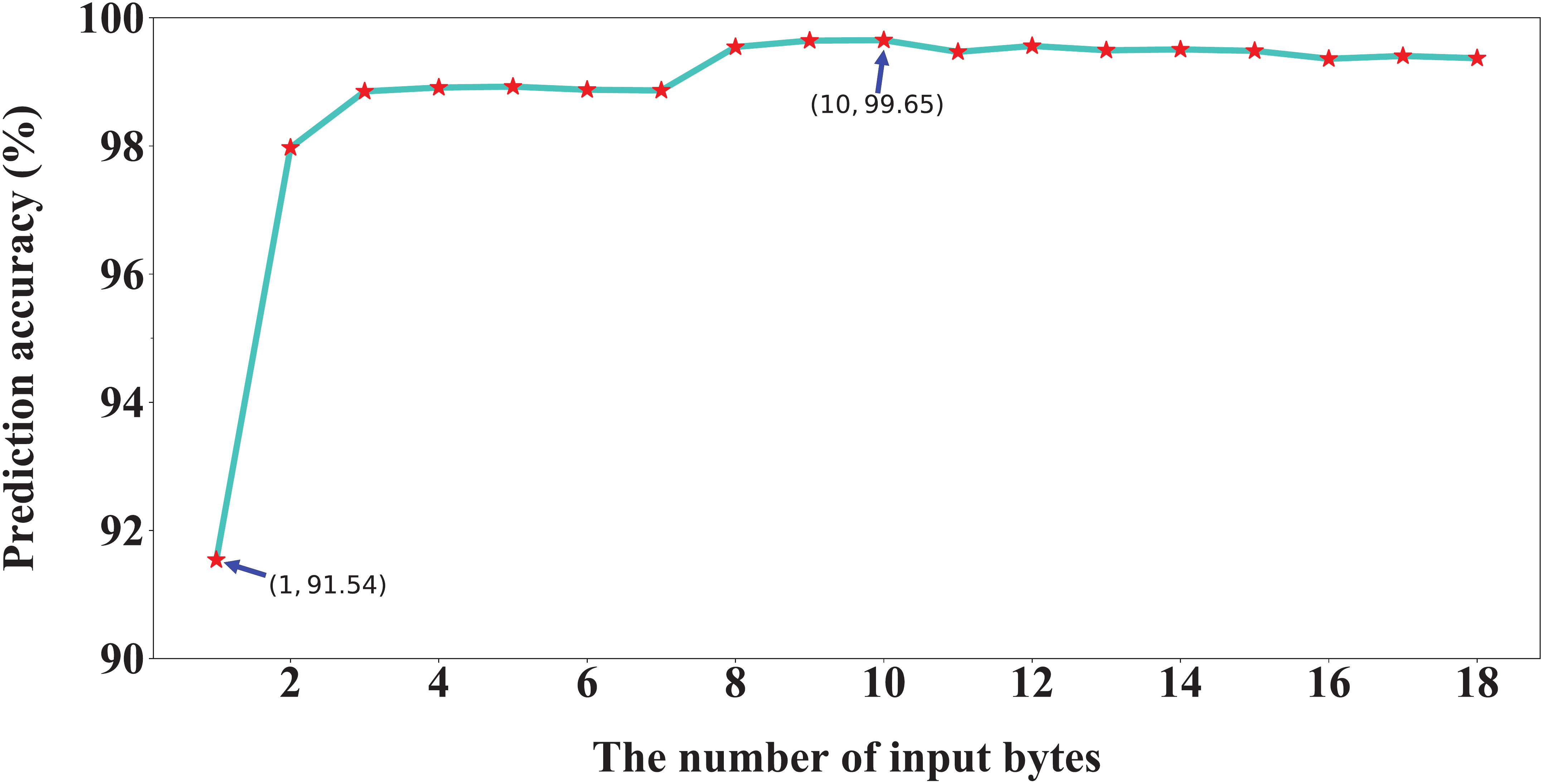}}
\caption{Modeling accuracies of RNN on VOLtA with different numbers of input elements using 10,000 CRPs.}
\label{Fig11}
\end{figure}

\subsubsection{ML Attacks VOLtA}

In the VOLtA, the current output of adder is related to the current input and the previous input. Therefore, the single input consists of multiple bytes, which is recorded as the input $X_t = \{x_{t-(m-1)}, ..., x_{t-2}, x_{t-1}, x_t\}$. The single output is 1-byte representing the current output of adder. Fig. \ref{Fig11} shows the modeling accuracies of RNN on VOLtA with different input bytes using 10,000 CRPs. We use the Hamming distance to evaluate the modeling accuracy. We can see from Fig. \ref{Fig11} that when $m = 1$, only the current input is used as the training input, the prediction accuracy of RNN is only 91.54$\%$. With the increasing of $m$, the modeling accuracy is further increased. The prediction accuracy reaches the highest 99.65$\%$ at $m = 10$. Therefore, we take $m = 10$ to conduct the following experiments.

\begin{figure}
\centerline{\includegraphics[width=\linewidth]{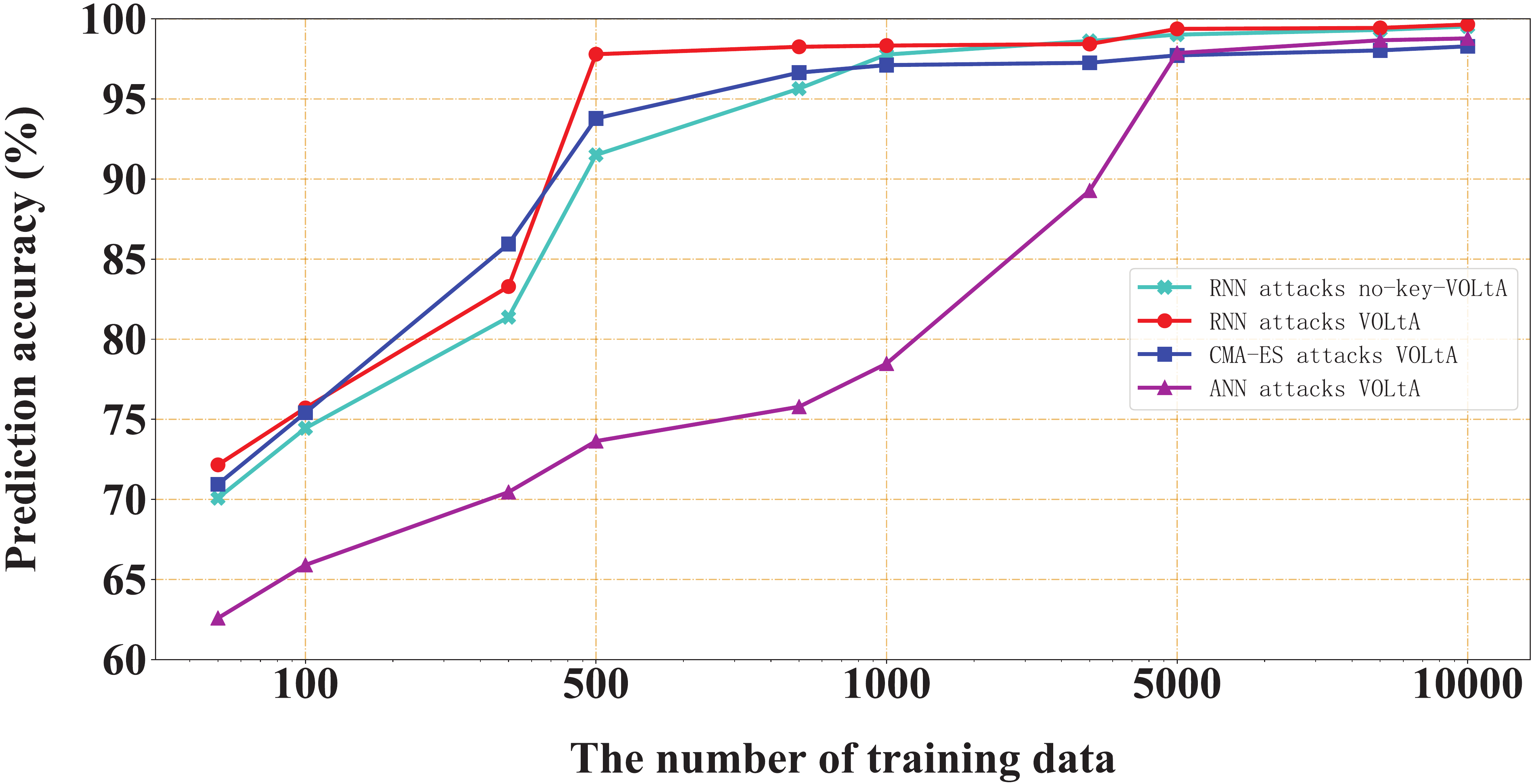}}
\caption{Modeling accuracies for VOLtA and no-key-VOLtA using 10,000 CRPs.}
\label{Fig12}
\end{figure}

The results of ML attacks on VOLtA and no-key-VOLtA are shown in Fig. \ref{Fig12}. When the RNN is used to attack the no-key-VOLtA, we collect two inputs of the adder as the challenge. When 500 CRPs are collected, the modeling accuracy of RNN model is more than 90$\%$; when 10,000 CRPs are collected, the prediction accuracy is up to 99.52$\%$. Therefore, the no-key-VOLtA is vulnerable to ML attacks. Next, we use ANN, RNN and CMA-ES to attack VOLtA. Since the output and one input have been obfuscated by the key in VOLtA, we only collect one input of adder as the challenge and the obfuscated output as the response. When 5,000 CRPs are collected, the modeling accuracies of ML attacks reach more than 95$\%$; when collecting 10,000 CRPs, the prediction accuracy of RNN is up to 99.65$\%$. Therefore, the modeling accuracy of RNN for VOLtA is just slightly higher than the no-key-VOLtA. In fact, the adder performs an approximate addition operation in VOS, where response $R = k_2 \oplus adder(C, X)$, if $X$ is an input, attackers can guess $k_2$ according to large amounts of $C$, $X$ and $R$; if $X$ is $k_1$, it will reduce the complexity of the model but increase the model security. Besides, the attackers need to collect vertical data to attack VOLtA, which requires to collect more data and consumes more time.

\begin{figure}
\centerline{\includegraphics[width=\linewidth]{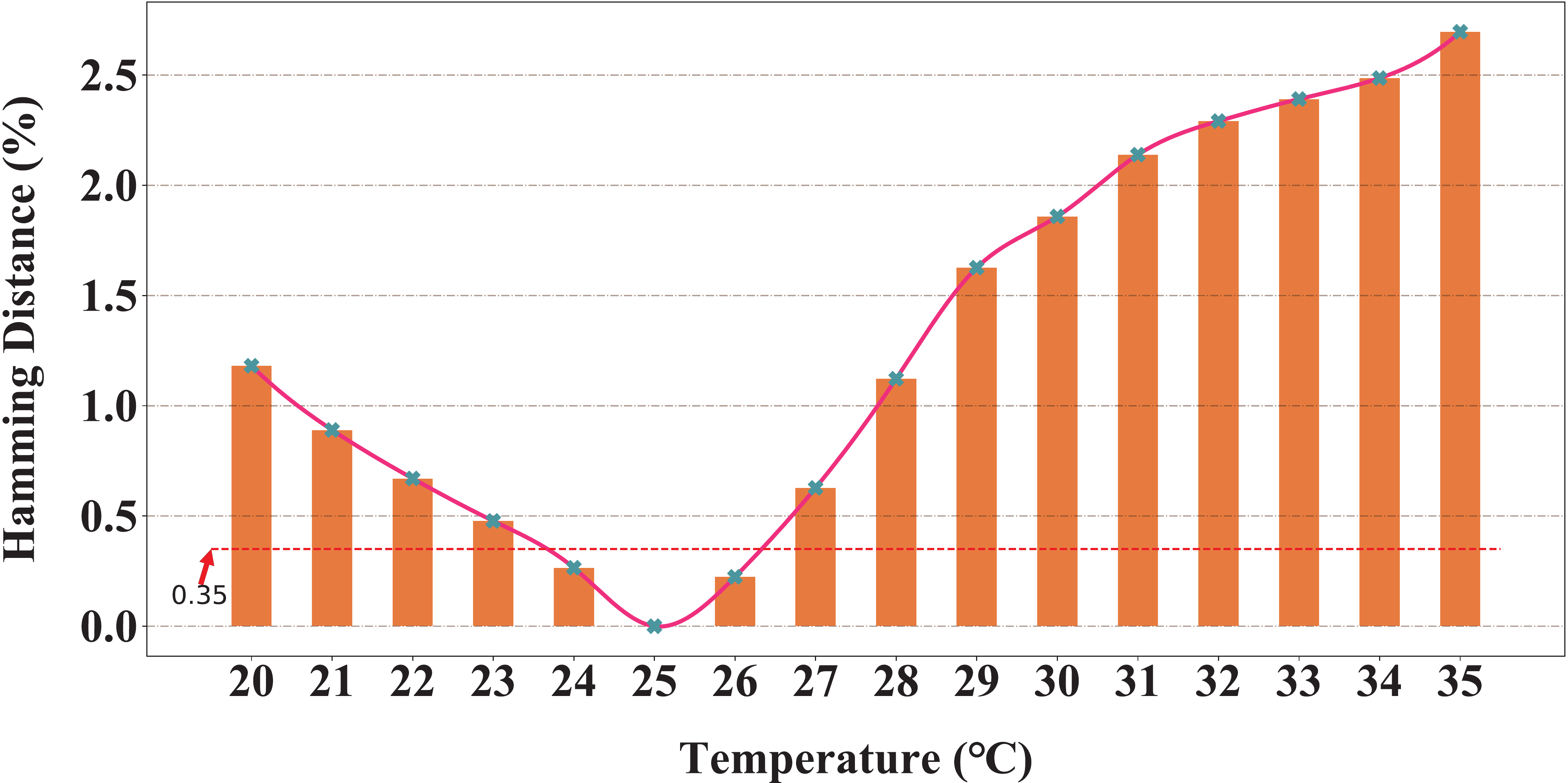}}
\caption{Reliability impacted by temperature variation (nominal temperature is 25$^{\circ}$C).}
\label{Fig13}
\end{figure}

\begin{figure}
\centering
\begin{minipage}[b]{.5\linewidth}
\centering
\includegraphics[width=.99\textwidth]{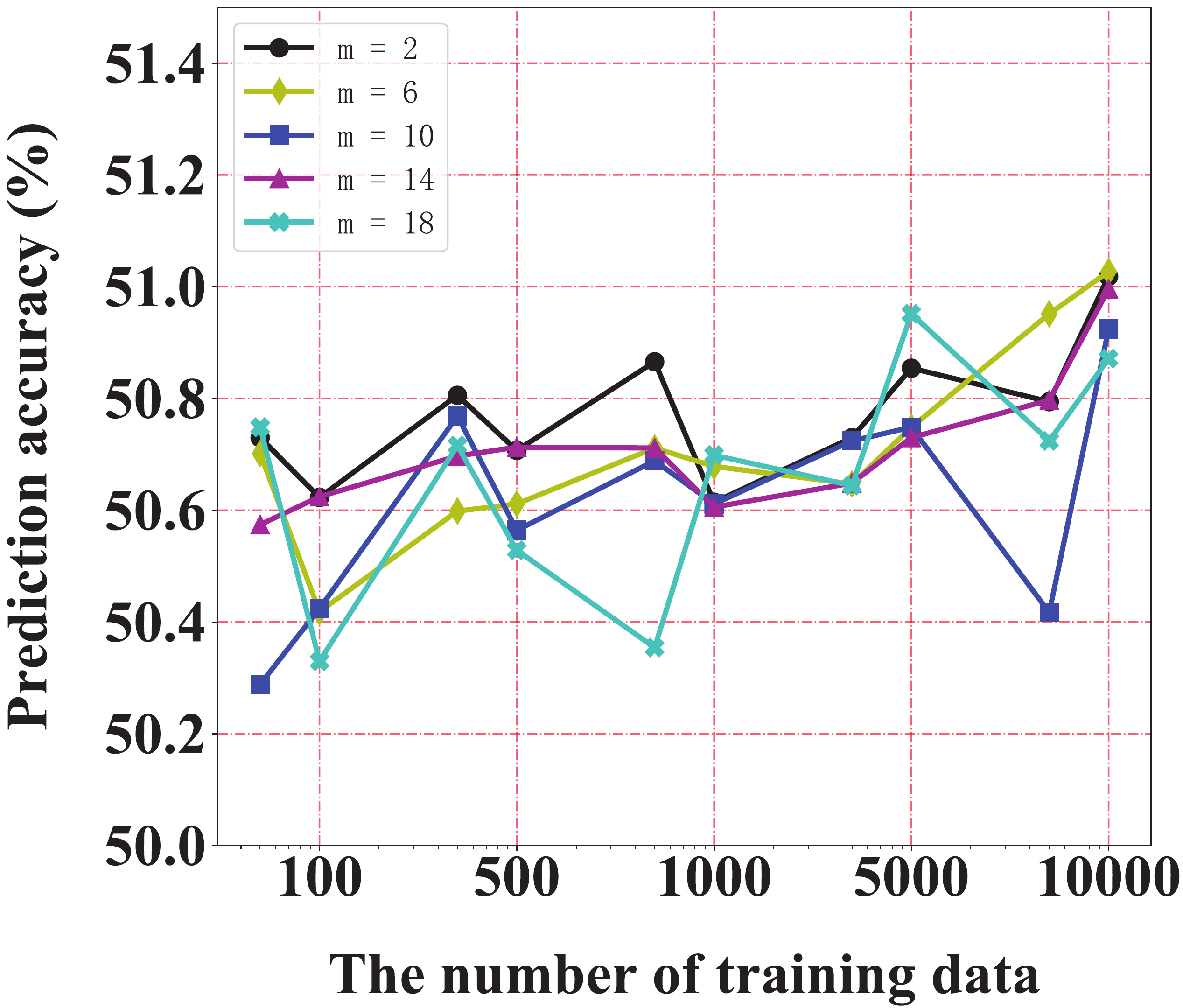}
\centerline{(a)}
\end{minipage}
\hspace{0cm}
\begin{minipage}[b]{.453\linewidth}
\centering
\includegraphics[width=.99\textwidth]{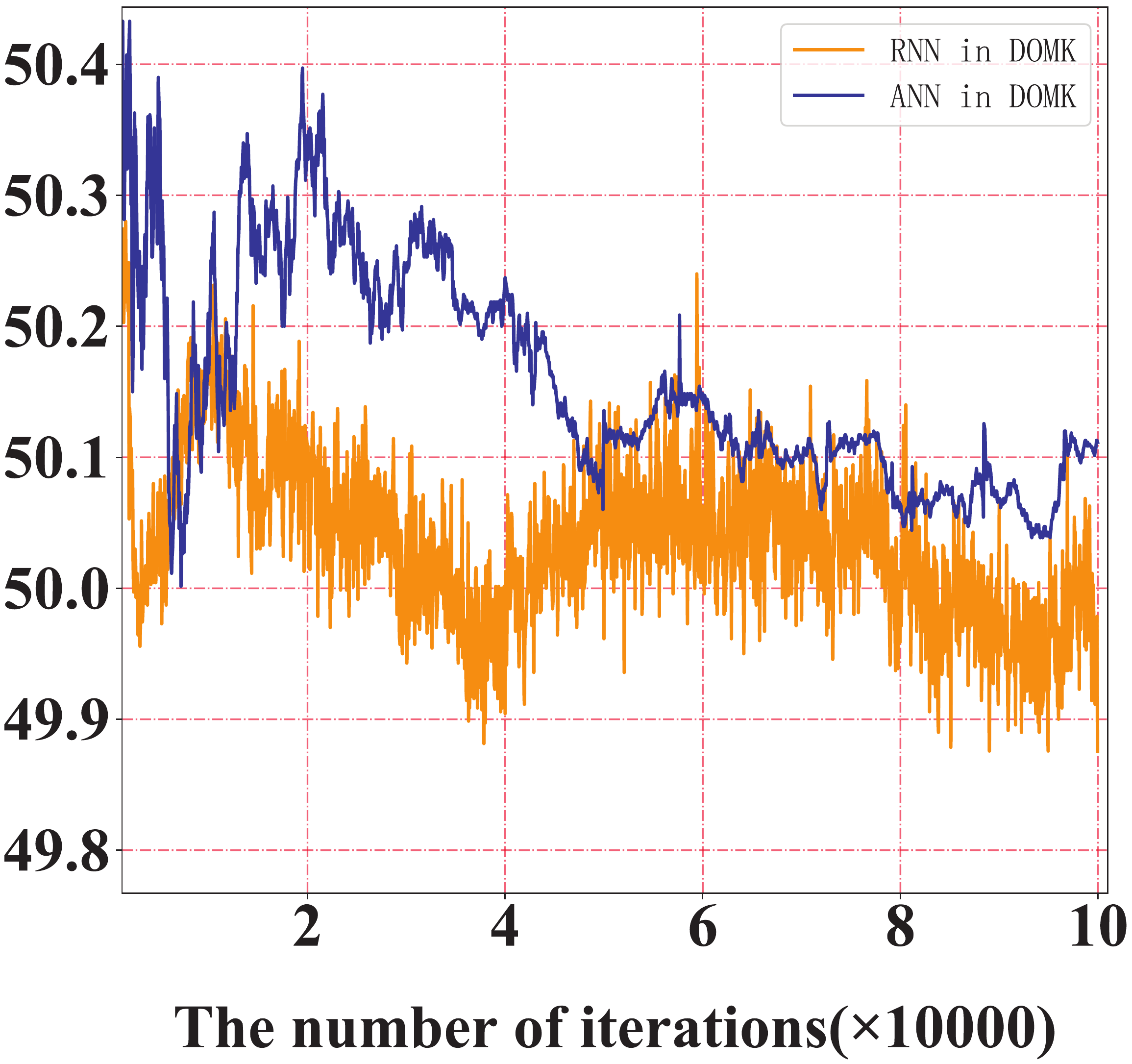}
\centerline{(b)}
\end{minipage}

\caption{The effectiveness of CSoS-based ML attacks resistant authentication.}
\label{Fig14}
\end{figure}

\subsubsection{VOLtA Reliability}

The intra Hamming distance (intra HD) of the responses is used to evaluate the reliability of VOLtA. We can see from Fig. \ref{Fig13} that the intra HD is around 0.47$\%$ when the temperature decreases from 25$^{\circ}$C to 23$^{\circ}$C, and it is about 0.62$\%$ when the temperature increases from 25$^{\circ}$C to 27$^{\circ}$C. The prediction accuracy of RNN is 99.65$\%$, while the error generated by the RNN is only 0.35$\%$ (see the red dotted line in Fig. \ref{Fig13}), which is less than the error caused by $\pm$2$^{\circ}$C. Unfortunately, the setting of threshold in VOLtA must consider the influence of temperature and other factors on the reliability. When the threshold is determined, the ML models can reach the threshold condition as well. Therefore, the VOLtA is vulnerable to ML modeling attacks.

%\subsection{Resistance to ML Attacks}

\subsection{Defenses}

\subsubsection{CSoS for VOLtA}

The effectiveness of CSoS-based ML attacks resistant authentication is evaluated. As shown in Fig. \ref{Fig14}(a), we set the input byte $m$ = 2, 6, 10, 14, 18; the training set is from 50 to 10,000. RNN is used to verify the effectiveness of the proposed protocol, in which the prediction accuracy selects the maximum during training. From the experimental results, we can see that even if the training set or $m$ is increased, the modeling accuracy is still between 50$\%$ and 51.2$\%$. The relationship between the iterations and the modeling accuracy of ML methods is shown in Fig. \ref{Fig14}(b). We can see that with the increasing of iterations, the prediction accuracies of ML methods are oscillating around 50.1$\%$. Therefore, the proposed CSoS-based authentication exhibits good resistance to learning-based attacks.

\subsubsection{CSoS for Arbiter PUF}

\begin{figure}
\centerline{\includegraphics[width=\linewidth]{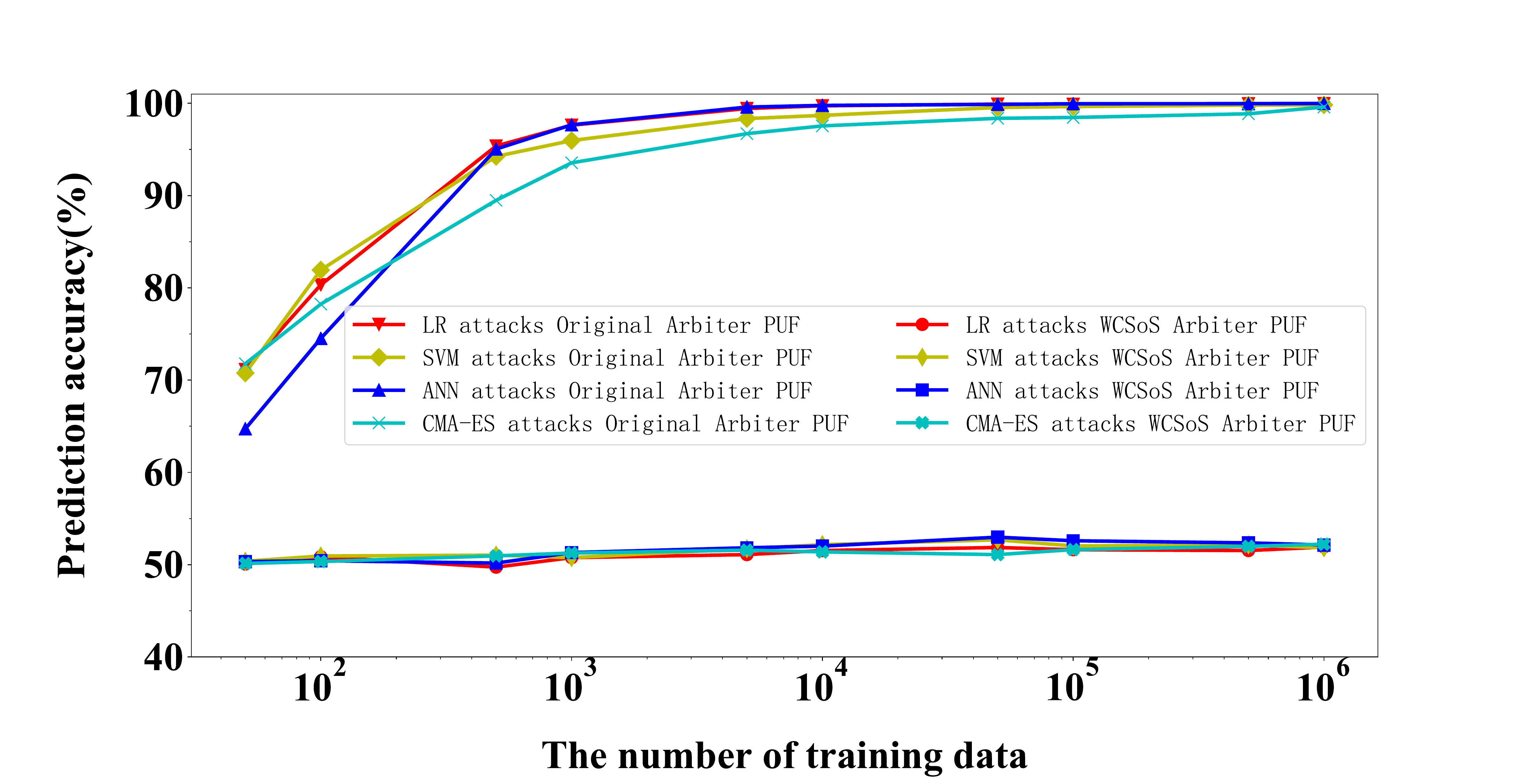}}
\caption{Modeling accuracies on the 64-bit Original and WCSoS Arbiter PUF using 1 million CRPs.}
\label{W_data}
\end{figure}

\begin{figure}
\centerline{\includegraphics[width=\linewidth]{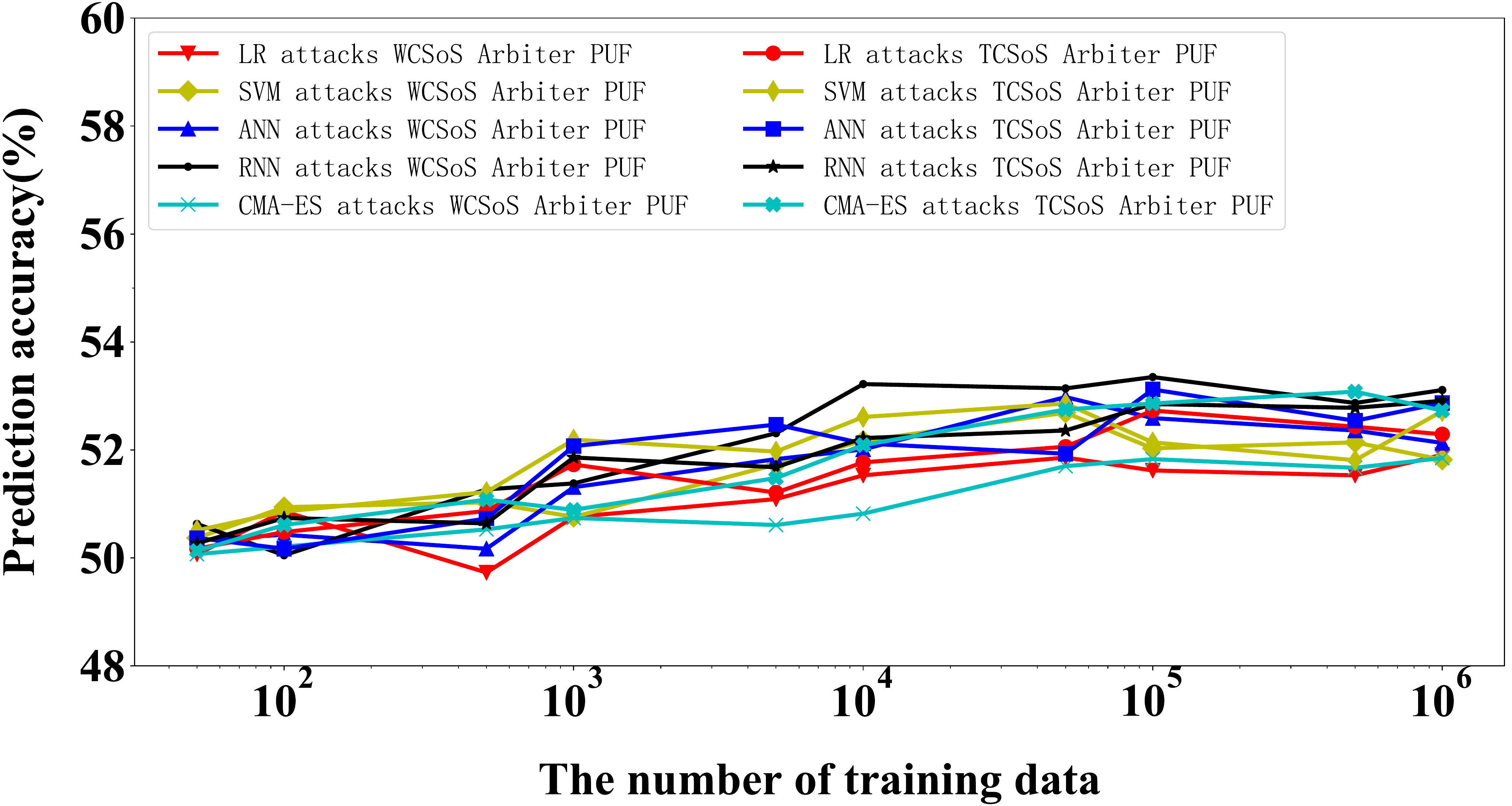}}
\caption{Modeling accuracies on the 64-bit WCSoS and TCSoS Arbiter PUF using 1 million CRPs.}
\label{T_data}
\end{figure}

%{\color{red} 实验还有很多数据没跑完，这里的图只是大致走势。}

Due to the limited number of CRPs that Hspice can collect, it is impossible to verify in VOLtA whether CSoS can still maintain high resistance to machine learning algorithms in larger data sets. For this reason, under a large data set for CSoS-based Arbiter PUF, we have evaluated the influence of ML attacks. We simulated $10^6$ CRPs to conduct this part of the experiment. R$\ddot{\rm u}$hrmair et al. \cite{Rührmair2013} demonstrates that modeling attacks can work both on simulated and silicon data, and the only difference is the case that the results on simulated data are noise free. However, by using more CRPs in the training stage, results from the real silicon could achieve the same accuracy rate (e.g., 99\%) compare to the simulated data. Furthermore, LR, SVM, ANN, RNN and CMA-ES are used to model WCSoS Arbiter PUF and TCSoS Arbiter PUF. The experimental results are shown in Fig. \ref{W_data} and Fig. \ref{T_data}.

\begin{figure}
\centerline{\includegraphics[width=\linewidth]{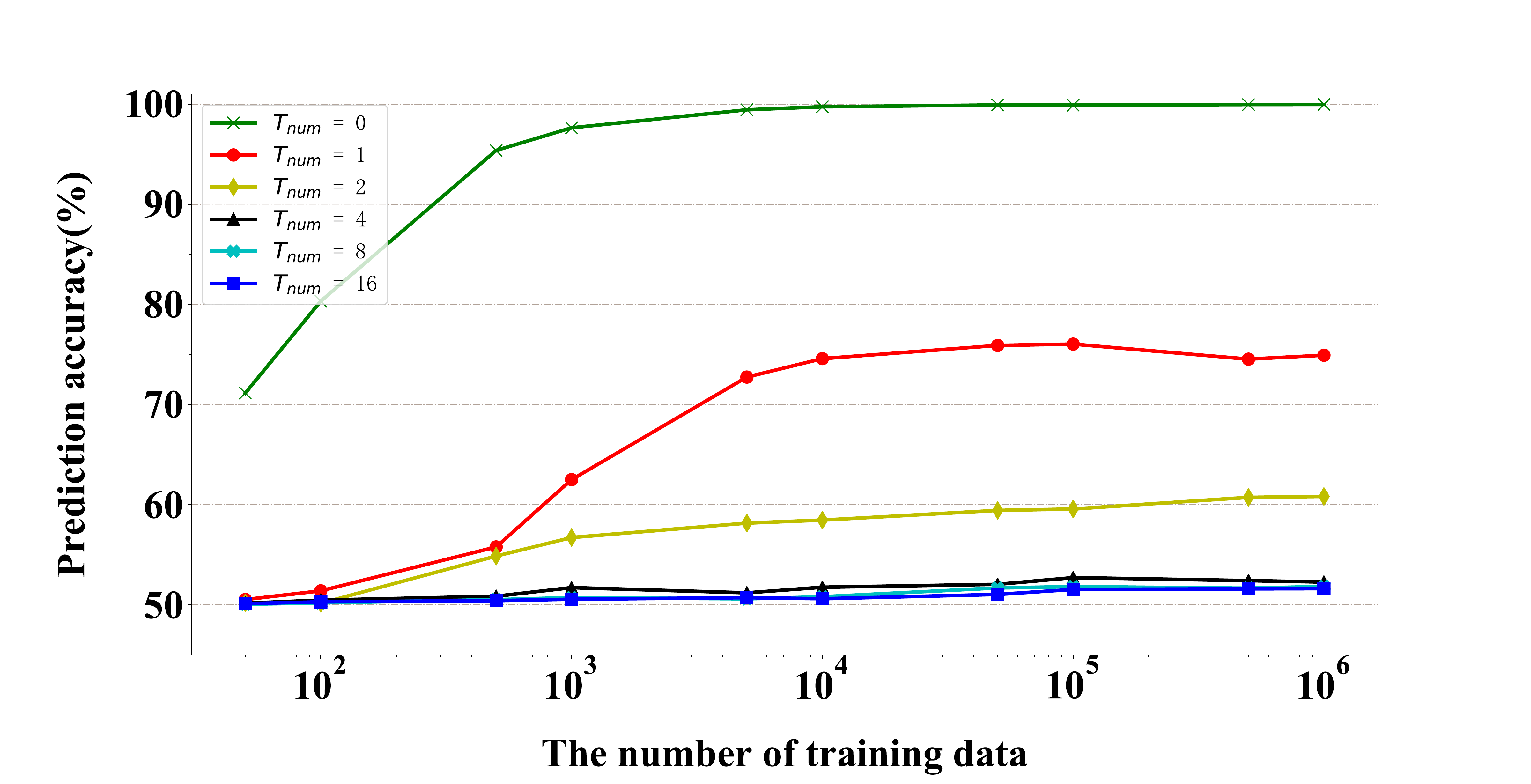}}
\caption{LR attack results on 64-bit TCSoS Arbiter PUF with different number of $T_{num}$ ($T_{num}$ is the bit number of TRNG).}
\label{T_num}
\end{figure}

\begin{figure}
\centerline{\includegraphics[width=\linewidth]{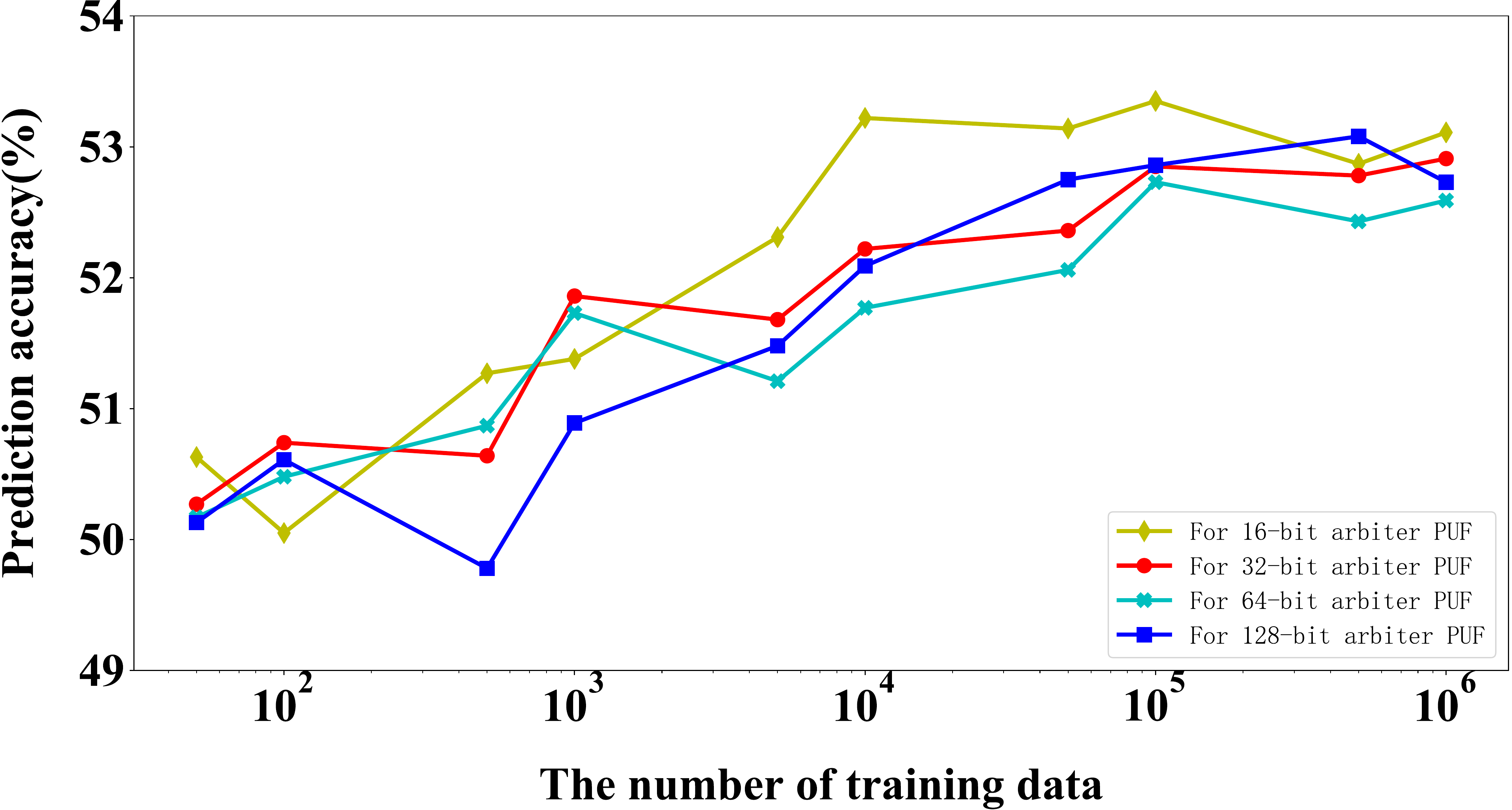}}
\caption{LR attack results on TCSoS Arbiter PUF with different number of $bit_{num}$ ($bit_{num}$ is the stage size of Arbiter PUF).}
\label{bit_num}
\end{figure}

%由于通过Hspice能够收集的CRPs的数量有限，无法在VOLtA中验证CSoS在更大数据集是否依旧能够对机器学习算法保持着高抵抗力，为此我们对CSoS-based Arbiter PUF 进行了大数据集的机器学习建模攻击。我们仿真生成了$10^6$ 对CRPs，并且使用LR，SVM，ANN，CMA-ES对WCSoS Arbiter PUF和TCSoS Arbiter PUF进行建模攻击。The experimental results are shown in Figs. \ref{W_data}, \ref{T_data}.

As shown in Fig. \ref{W_data}, we use ML to attack Arbiter PUF without deploying the obfuscation mechanism. When 5,000 CRPs are collected, the modeling accuracies of ML algorithms are more than 95\%; When $10^6$ CRPs are collected, LR can achieve 99.87\% modeling accuracy. Obviously, the Arbiter PUF without deploying the defense mechanism can be broken by ML algorithms easily. When ML methods are utilized to model WCSoS Arbiter PUF, the modeling accuracy did not increase significantly as the training set growing. Even if $10^6$ CRPs are collected, the accuracy is still below 54\%, which shows that CSoS still maintains good anti-modeling ability under the massive data set. In Fig. \ref{T_data}, we compare the WCSoS Arbiter PUF and TCSos Arbiter PUF modeling attacks, where a 4-bit TRNG is used in TCSoS. Experimental results that both TCSoS and WCSoS show good resistance to ML attacks.

%如Fig. \ref{W_data}所示，我们使用机器学习对未部署任何混淆机制的Arbiter PUF进行攻击，当收集到5,000 CRPs时，机器学习算法的建模准确率都在95\%以上；当收集到$10^6$ CRPs 时，LR 可以达到99.87\% 建模准确率。显然，未部署任何防御机制的Arbiter PUF可以被机器学习算法轻易攻破。在使用机器学习对WCSoS Arbiter PUF 建模时，随着训练集的增长，建模准确率没有明显的增长。即使收集到$10^6$ CRPs，准确率依旧在54\%以下，可见CSoS 在大数据集下依旧保持着良好的抗建模能力。在Fig. \ref{T_data} 中，我们对WCSoS Arbiter PUF和TCSoS Arbiter PUF建模攻击进行比较，其中TCSoS中采用4位TRNG。从实验结果可以看出，TCSoS的抗机器学习建模能力要略强于WCSoS。

TCSoS has high flexibility to deploy different levels of TRNG based on its own security requirements and affordable computing power. As shown in Fig. \ref{T_num}, we use LR to model the 64-bit TCSoS Arbiter PUF with TRNG bits $T_{num} = 0, 1, 2, 4, 8, 16$ ($T_{num} = 0$ means TCSoS is not deployed), when $T_{num}$ = 1 and 2, the modeling accuracy of LR can reach 74.93\% and 60.83\% respectively, which does not meet the security requirements; when $T_{num}$ = 4 and 8, even if collecting $10^6$ CRPs, the modeling accuracy of LR is still below 54\%. It is worth mentioning that when $T_{num} = 16$, LR only has a modeling accuracy of 51.63\%. However, the computational cost of server authentication at this time will be $2^{16} = 65,536$ times more than normal conditions. Therefore, $T_{num} = 4$ or $8$ is the empirical value we recommended in the actual deployment.

Next, we verify the effectiveness of TCSoS ($T_{num} = 4$) for Arbiter PUF with different stage sizes. As shown in Fig. \ref{bit_num}, regardless of the stage size of Arbiter PUF, the modeling accuracy of ML for TCSoS has been reduced and finally stabilized around 54\%. Hence, TCSoS provides good obfuscation ability for Arbiter PUFs with different stage sizes. Moreover, we also verify the effectiveness of different ML algorithms on TCSoS-based Arbiter PUF with different $T_{num}$ and $bit_{num}$ (stage size). Table II gives detail experimental data which demonstrate that TCSoS can effectively obfuscate the mapping relationship of CRPs in Arbiter PUF with different stage sizes and shows good resistance to several ML attack methods.

%从实验结果可以看出，不管stage size 是多少，机器学习对TCSoS的建模准确率虽然在开始都有所提升但是最后都在54\%附近趋于平稳，可见TCSoS对不同stage size的PUF 都能起到很好的混淆效果。Furthermore，我们还验证了不同机器学习算法对不同T_num和stage size 下的TCSoS Arbiter PUF的抗建模效果，实验数据如表二所示。

%TCSoS 拥有较高的灵活性，可以根据自身的安全性需求以及可承受的运算能力选择不同位数的TRNG 进行部署。如Fig. \ref{T_num}所示，我们采用LR对TRNG的位数 $T_{num} = 1, 2, 4, 8, 16$ 的64-bit TCSoS Arbiter PUF 进行建模攻击，当 $T_{num} = 1,2$ 时，LR 的建模准确率分别可以达到72.82\%和57.73\%，显然还不足以满足安全性需求；当$T_{num} = 4,8$ 时，即使收集到$10^6$ CRPs, LR的建模准确率依旧在54\%以下；值得一提的是，在 $ T_{num} = 16$ 时，LR仅有50.86\%的建模准确率，然而此时服务器认证的计算开销将为正常情况下的 $2^{16} = 65,536$ 倍。因此，我们建议$T_{num} = 4 or 8$ 为最佳。

%在Fig. \ref{bit_num}，我们进行了LR attack results on different TCSoS Arbiter PUF。

%在表二中，进行了Modeling accuracies on different TCSoS Arbiter PUF with different number of $T_{num}$。

\begin{table}
\caption{Modeling accuracies on TCSoS Arbiter PUF with different number of $T_{num}$ and $bit_{num}$ using $10^5$ CRPs.}
\centering
\begin{tabular}{*{6}{c}}
\toprule
\multirow{2}{*}{$T_{num}$} & \multirow{2}{*}{MLs}  & \multicolumn{4}{c}{$Bit_{num}$}  \\
                \cmidrule(r){3-6}
                & & 16 & 32 & 64 & 128 \\

\midrule
\multirow{4}{*}{0} & LR   &99.99\% &99.99\% &99.89\% &99.95\% \\
                    {}& SVM  &99.99\% &99.97\% &99.83\% &99.89\% \\
                    {}& ANN  &99.99\% &99.99\% &99.96\% &99.98\% \\
                    {}& CMA-ES  &99.99\% &99.99\% &99.99\% &99.99\% \\
\cmidrule(r){1-6}
\multirow{5}{*}{1} & LR   &75.06\% &73.23\% &76.04\% &74.77\% \\
                    {}& SVM  &71.87\% &74.32\% &72.39\% &75.25\% \\
                    {}& ANN  &78.83\% &72.06\% &71.67\% &80.54\% \\
                    {}& RNN  &76.96\% &77.23\% &76.79\% &73.36\% \\
                    {}& CMA-ES  &79.36\% &76.07\% &78.26\% &73.60\% \\
\cmidrule(r){1-6}
\multirow{5}{*}{2} & LR  &59.96\% &61.23\% &59.58\% &63.03\% \\
                    {}& SVM  &56.51\% &66.32\% &58.11\% &64.05\% \\
                    {}& ANN  &61.33\% &58.74\% &57.70\% &60.22\% \\
                    {}& RNN  &58.73\% &64.54\% &61.96\% &67.14\% \\
                    {}& CMA-ES  &58.28\% &56.23\% &63.31\% &62.26\% \\
\cmidrule(r){1-6}
\multirow{5}{*}{4} & LR   &53.35\% &52.85\% &52.73\% &52.86\% \\
                    {}& SVM  &52.05\% &51.81\% &52.14\% &52.23\% \\
                    {}& ANN  &53.07\% &53.16\% &53.12\% &51.99\% \\
                    {}& RNN  &52.18\% &51.87\% &51.59\% &52.83\% \\
                    {}& CMA-ES  &53.06\% &52.30\% &51.96\% &52.30\% \\
\cmidrule(r){1-6}
\multirow{5}{*}{8} & LR   &51.73\% &52.49\% &51.83\% &52.38\% \\
                    {}& SVM  &51.57\% &52.81\% &51.77\% &52.49\% \\
                    {}& ANN  &52.13\% &52.16\% &51.69\% &51.37\% \\
                    {}& RNN  &52.13\% &51.80\% &51.97\% &52.43\% \\
                    {}& CMA-ES  &52.09\% &51.27\% &52.36\% &51.97\% \\
\cmidrule(r){1-6}
\multirow{5}{*}{16} & LR   &51.99\% &52.37\% &51.55\% &51.69\% \\
                    {}& SVM  &52.26\% &51.59\% &51.72\% &52.50\% \\
                    {}& ANN  &51.87\% &52.32\% &51.92\% &52.42\% \\
                    {}& RNN  &52.32\% &52.29\% &51.87\% &52.69\% \\
                    {}& CMA-ES  &52.28\% &51.98\% &52.31\% &51.72\% \\
\bottomrule
\end{tabular}
\end{table}

\section{Conclusion}
In this paper, we have reevaluated the security of the VOS-based authentication protocol and implemented several high-accuracy ML modeling attacks on VOLtA. Experimental results show that the VOLtA is vulnerable to ML attacks, and the prediction accuracy of RNN is up to 99.65$\%$. To resist the ML attacks on the VOLtA, this paper proposes a novel challenge self-obfuscation structure (CSoS), which lowers the prediction accuracy of ML on the VOLtA to 51.2$\%$. Furthermore, our proposed CSoS exhibits good obfuscation ability for both VOLtA and strong PUFs. We collect $10^6$ CRPs of a Arbiter PUF deployed with CSoS and modeled it using LR, SVM, ANN, RNN and CMA-ES. The experimental results show that modeling accuracy is reduced to 54$\%$.

\vfill


\begin{thebibliography}{1}
\bibitem{IoT}
Wikipedia, ''Internet of things,'' [Online]. Available: https://en.wikipedia.org/wiki/Internet$\_$of$\_$things
%1
%\bibitem{Li2015}
%S. Li, L. D. Xu, and S. Zhao, ``The internet of things: a survey,'' \emph{Inf. Syst. Front.}, vol. 17, no. 2, pp. 243-259, Apr. 2015.
%2
%\bibitem{Han2013a}
%C. Han, J. M. Jornet, E. Fadel, and I. F. Akyildiz, ``A cross-layer communication module for the Internet of Things,'' \emph{Comput. Networks}, vol. 57, no. 3, pp. 622-633, Feb. 2013.
%%3
%\bibitem{Zhao2013}
%K. Zhao and L. Ge, ``A Survey on the Internet of Things Security,'' \emph{in 2013 Ninth International Conference on Computational Intelligence and Security}, pp. 663-667, 2013.
%4
\bibitem{IoT2017}
``Smart Summit Asia: Identifying Key Technology Drivers for Wider Adoption of Connected Solutions,'' [Online]. Available:
https://technology.ihs.com/587648, 2017.
%5
\bibitem{Woolf2016}
``DDoS attack that disrupted internet was largest of its kind in history, experts say,'' [Online]. Available: https://www.theguardian.com/technology/2016/oct/26/ddos-attack-dyn-mirai-botnet, 2016.
%6
\bibitem{Antonakakis2017}
M. Antonakakis et al., ``Understanding the Mirai Botnet This paper is included in the Proceedings of the Understanding the Mirai Botnet,'' \emph{USENIX Secur.}, 2017.
%7
\bibitem{Ruhrmair2014}
U. R$\ddot{\rm u}$hrmair and D. E. Holcomb, ``PUFs at a glance,'' in \emph{Design, Automation and Test in Europe (DATE)}, 2014, pp. 1-6.
%8
\bibitem{Zhang2014}
J. L. Zhang, G. Qu, Y. Q. Lv, and Q. Zhou, ``A survey on silicon PUFs and recent advances in ring oscillator PUFs,'' \emph{J. Comput. Sci. Technol.}, vol. 29, no. 4, pp. 664-678, 2014.
%9
\bibitem{Arafin2017}	
M. T. Arafin, M. Gao, and G. Qu, ``VOLtA: Voltage Over-scaling Based Lightweight Authentication for IoT Applications,'' \emph{2017 22nd Asia and South Pacific Design Automation Conference (ASP-DAC). IEEE}, pp. 336-341, 2017.
%10

\bibitem{Zhang2018}
H. Su and J. Zhang, ``Machine Learning Attacks on Voltage Over-scaling-based Lightweight Authentication,'' \emph{Asian Hardware Oriented Security and Trust Symposium}, 2018.

\bibitem{Pappu2002}	
R. Pappu, B. Recht, J. Taylor, N. Gershenfeld, ``Physical one-way functions,'' \emph{Science}, vol. 297, no.5589, pp.2026-2030, Sep. 2002.

%\bibitem{Hammouri2008}
%G. Hammouri, E. $\ddot{\rm O}$zt$\ddot{\rm u}$rk, and B. Sunar, ``A tamper-proof and lightweight authentication scheme,'' \emph{Pervasive Mob. Comput.}, vol. 4, no. 6, pp. 807-818, Dec. 2008.
%11
\bibitem{Lee2004}
J. W. Lee, Daihyun Lim, B. Gassend, G. E. Suh, M. van Dijk, and S. Devadas, ``A technique to build a secret key in integrated circuits for identification and authentication applications,'' in \emph{2004
Symposium on VLSI Circuits.} Digest of Technical Papers (IEEE Cat. No.04CH37525), pp. 176-179.

\bibitem{Lim2005}
D. Lim, J. W. Lee, B. Gassend, G. E. Suh, M. van Dijk, and S. Devadas, ``Extracting Keys from Integrated Circuits,'' \emph{IEEE Trans. Very Large Scale Integr. Syst.}, vol. 13, no. 10, pp. 1200-1205, 2005.
%12
\bibitem{Vijayakumar2015}
A. Vijayakumar and S. Kundu, ``A Novel Modeling Attack Resistant PUF Design based on Non-linear Voltage Transfer Characteristics,'' in \emph{Design, Automation And Test in Europe (DATE)}, 2015, 2015, pp. 653-658.
%13
\bibitem{Majzoobi2008a}
M. Majzoobi, F. Koushanfar, and M. Potkonjak, ``Lightweight secure PUFs,'' \emph{IEEE/ACM Int. Conf. Comput. Des. Dig. Tech. Pap. ICCAD}, vol. 1, no. 1, pp. 670-673, 2008.
%14
\bibitem{Sahoo2014}
D. P. Sahoo, S. Saha, D. Mukhopadhyay, R. S. Chakraborty, and H. Kapoor, ``Composite PUF: A new design paradigm for Physically Unclonable Functions on FPGA,'' \emph{Proc. IEEE Int. Symp. Hardware-Oriented Secur. Trust. HOST 2014}, pp. 50-55, 2014.
%15
\bibitem{Holcomb2007}	
D. E. Holcomb, W. P. Burleson, and K. Fu, ``Initial SRAM state as a fingerprint and source of true random numbers for RFID tags,'' \emph{Proc. Conf. RFID Secur.}, vol. 58, no. 9, pp. 1-12, 2007.
%16
\bibitem{Suh2007}
G. E. Suh and S. Devadas, ``Physical Unclonable Functions for Device Authentication and Secret Key Generation,'' in \emph{44th ACM/IEEE Design Automation Conference}, pp.9-14, 2007.
%17
\bibitem{Tuyls2006}
P. Tuyls, G.-J. Schrijen, B. skoric, J. van Geloven, N. Verhaegh, and R. Wolters, ``Read-Proof Hardware from Protective Coatings,'' \emph{Cryptogr. Hardw. Embed. Syst.}, pp. 369-383, 2006.
%18
\bibitem{Sauer2017}
M. Sauer, P. Raiola, L. Feiten, B. Becker, U. R$\ddot{\rm u}$hrmair, and I. Polian, ``Sensitized path PUF: A lightweight embedded physical unclonable function,'' \emph{Proc. Des. Autom. Test Eur.}, pp. 680-685, 2017.
%19
\bibitem{Zhang2018}
J. Zhang and L. Wan, ``CMOS: Dynamic Multi-key Obfuscation Structure for Strong PUFs,'' 2018.
%20
\bibitem{Venkatesan2011}
R. Venkatesan, A. Agarwal, K. Roy, and A. Raghunathan, ``MACACO: Modeling and analysis of circuits for approximate computing,'' \emph{in IEEE/ACM International Conference on Computer-Aided Design, Digest of Technical Papers, ICCAD}, pp. 667-673, 2011.
%21
\bibitem{Chen2013}
J. N. Chen and J. H. Hu, ``Energy-Efficient Digital Signal Processing via Voltage-Overscaling-Based Residue Number System,'' \emph{IEEE Trans. Very Large Scale Integr. Syst.}, vol. 21, no. 7, pp. 1322-1332, Jul. 2013.
%22
\bibitem{Rokhani2006}
F. Z. Rokhani and G. E. Sobelman, ``Low-power bus transform coding for multilevel signals,'' \emph{in IEEE Asia-Pacific Conference on Circuits and Systems, Proceedings, APCCAS}, 2006, pp. 1272-1275.

%23
\bibitem{Gutnik1997}
V. Gutnik and A. P. Chandrakasan, ``Embedded power supply for low-power DSP,'' \emph{IEEE Trans. Very Large Scale Integr. Syst.}, vol. 5, no. 4, pp. 425-435, Dec. 1997.
%24
\bibitem{Chabini2004}
N. Chabini and W. Wolf, ``Reducing dynamic power consumption in synchronous sequential digital designs using retiming and supply voltage scaling,'' \emph{IEEE Trans. Very Large Scale Integr. Syst.}, vol. 12, no. 6, pp. 573-589, Jun. 2004.
%25
\bibitem{Liu2011}
R. Liu and K. K. Parhi, ``Power reduction in frequency-selective FIR filters under voltage overscaling,'' \emph{IEEE J. Emerg. Sel. Top. Circuits Syst.}, vol. 1, no. 3, pp. 343-356, 2011.
%26
\bibitem{Han2013b}
J. Han and M. Orshansky, ``Approximate Computing: An Emerging Paradigm For Energy-Efficient Design,'' \emph{IEEE Test Symposium}, vol. 370, pp. 1-6, 2013.
%27
\bibitem{Li2013}
H. Li, J. Hu, and J. Chen, ``A novel low-power filter design via reduced-precision redundancy for voltage overscaling applications,'' \emph{IEEE Glob. Telecommun. Conf.}, pp. 3282-3287, 2013.
%28
\bibitem{Rührmair2010}
U. R$\ddot{\rm u}$hrmair, F. Sehnke, J. Selter, G. Dror, S. Devadas, and J. Schmidhuber, ``Modeling attacks on physical unclonable functions,'' \emph{Proc. 17th ACM Conf. Comput. Commun. Secur}, p. 237-249, 2010.
%29
\bibitem{Bishop2015}
C. M. Bishop, ``Pattern Recognition and Machine Learning'',\emph{Information Science and Statistics, Springer-Verlag New York, Inc.}, pp.049901, 2006.
%30
\bibitem{Hornik1991}
K. Hornik, ``Approximation capabilities of multilayer feedforward networks,'' \emph{Neural Networks}, vol. 4, no. 2, pp. 251-257, 1991.
%31
\bibitem{Rosenblatt1957}
F. Rosenblatt, ``The Perceptron - A Perceiving and Recognizing Automaton,'' \emph{Math. Stat}, 1957.
%32
\bibitem{Hansen2001}
N. Hansen and A. Ostermeier, ``Completely Derandomized Self-Adaptation in Evolution Strategies,'' \emph{Evol. Comput.}, vol. 9, no. 2, pp. 159-195, Jun. 2001.
%33
\bibitem{Hansen2016}
N. Hansen, ``The CMA Evolution Strategy: A Tutorial,'' 2016.
%34
\bibitem{Pang2017}
Z. H. Pang, J. Zhang, Q. Zhou, S. Q. Gong, X. Qian and B. Tang, ``Crossover Ring Oscillator PUF,'' \emph{in 2017 18th International Symposium on Quality Electronic Design (ISQED)}, pp. 237-243, 2017.
%35
\bibitem{Ma2018}
Q. Ma, C. Gu, N. Hanley, C. Wang, W. Liu, and M. O’Neill, ``A machine learning attack resistant multi-PUF design on FPGA,'' in \emph{2018 23rd Asia and South Pacific Design Automation Conference (ASP-DAC).}, 2018, pp. 97-104.
%36

%37
\bibitem{Ye2017}
J. Ye, Y. Hu, and X. Li, ``VPUF: Voter based physical unclonable function with high reliability and modeling attack resistance,'' in \emph{2017 IEEE 23rd International Symposium on On-Line Testing and Robust System Design (IOLTS)}, 2017, pp. 74-79.

\bibitem{Kocher1996}
P. C. Kocher, ``Timing Attacks on Implementations of Diffie-Hellman, RSA, DSS, and Other Systems,'' 1996, pp. 104-113.

\bibitem{Merli2011a}
D. Merli, D. Schuster, F. Stumpf, and G. Sigl, ``Side-Channel Analysis of PUFs and Fuzzy Extractors,'' in \emph{International Conference on Trust and Trustworthy Computing}, 2011, pp. 33-47.

\bibitem{Merli2013}
D. Merli, J. Heyszl, B. Heinz, D. Schuster, F. Stumpf, and G. Sigl, ``Localized electromagnetic analysis of RO PUFs,'' in \emph{IEEE International Symposium on Hardware-Oriented Security and Trust (HOST)}, 2013, pp. 19-24.

%38
\bibitem{Stine2007}
J. E. Stine et al., ``FreePDK: An Open-Source Variation-Aware Design Kit,'' \emph{IEEE International Conference on Microelectronic Systems Education}, pp. 173-174, 2007.
%39
\bibitem{Rührmair2013}
U. R$\ddot{\rm u}$hrmair et al., ``PUF Modeling Attacks on Simulated and Silicon Data,'' \emph{IEEE Trans. Inf. Forensics Secur.}, vol. 8, no. 11, pp. 1876-1891, Nov. 2013.

\end{thebibliography}
\end{document}